\documentclass[12pt,preprint]{aastex}
\usepackage{emulateapj5}

\newcommand{\kms}{km\,s$^{-1}$} 
\newcommand{\thco}{$^{13}$CO}   
\newcommand{\ceo}{C$^{18}$O}   
\newcommand{\hts}{H$_2$S}
\newcommand{\jyb}{Jy\,beam$^{-1}$}

\newcommand{\wgm}{II}

\newcommand{\HII}{H\,{\sc ii}}
\newcommand{\too}{$\rightarrow$}

\begin{document}

\title{High-velocity gas toward hot molecular cores: evidence for
  collimated outflows from embedded sources}  

\author{A.G. Gibb} 
\affil{Department of Astronomy, University of Maryland,
College Park, MD 20742, USA}
\email{agg@astro.umd.edu}

\author{F. Wyrowski}
\affil{Max-Planck-Institut f\"ur Radioastronomie, Auf dem
H\"ugel 69, D-53121 Bonn, Germany}
\email{wyrowski@mpifr-bonn.mpg.de}

\author{L.G. Mundy}
\affil{Department of Astronomy, University of Maryland,
College Park, MD 20742, USA}
\email{lgm@astro.umd.edu}

\begin{abstract} 
 
We present observations made with the Berkeley-Illinois-Maryland
Association millimeter array of the H$_2$S 2(2,0)$\rightarrow$ 2(1,1)
and C$^{18}$O 2$\rightarrow$ 1 transitions toward a sample of four hot
molecular cores associated with ultracompact H\,{\sc ii} regions:
G9.62+0.19, G10.47+0.03, G29.96$-$0.02 and G31.41+0.31. The angular
resolution varies from 1.5 to 2.4 arcsec, corresponding to scales of
$\sim$0.06 pc at the distance of these sources. High-velocity wings
characteristic of molecular outflows are detected toward all four
sources in the H$_2$S line. In two cases (G29.96 and G31.41) red- and
blueshifted lobes are clearly defined and spatially separate,
indicating that the flows are collimated. We also confirm the previous
detection of the outflow in G9.62F. Although the gas-phase H$_2$S
abundance is not well constrained, assuming a value of $10^{-7}$
yields lower limits to total outflow masses of $\sim$8 M$_\odot$,
values which are consistent with the driving sources being massive
protostars. Linear velocity gradients are detected in both C$^{18}$O
and H$_2$S across G9.62, G29.96 and, to a lesser extent, G31.41. These
gradients are observed to be at a different position angle to the
outflow in G9.62F and G29.96, suggestive of a rotation signature in
these two hot cores. Our observations show that these hot cores
contain embedded massive protostellar objects which are driving
bipolar outflows. Furthermore, the lack of strong centimeter-wave
emission toward the outflow centers in G29.96 and G31.41 indicates
that the outflow phase begins prior to the formation of a detectable
ultracompact H\,{\sc ii} region.

\end{abstract}


\section{Introduction}

Hot molecular cores are compact, dense cores of gas which display a
rich array of chemical species and are frequently found in association
with ultracompact \HII\ regions (e.g.\ Kurtz et al.\ 2000).
Temperatures in the region of 100--300 K have been derived from
ammonia and methyl cyanide observations (Cesaroni et al.\ 1994, Olmi
et al.\ 1996). In addition to the high temperature a striking feature
of hot cores is the elevated abundance of saturated molecules and rare
species compared with dark clouds such as TMC-1, which chemical models
show can only arise if chemical reactions on grain surfaces are
included (Brown, Millar \& Charnley 1988).  However, it has been
unclear as to whether all hot cores marked the actual sites of massive
star formation or if some were remnant clumps heated by nearby OB
stars (e.g.\ Watt \& Mundy 1999).

Many studies have been made of hot cores, and observations with
interferometers have begun to reveal the geometry of the gas and dust
in these regions (Blake et al.\ 1996; Wright, Plambeck \& Wilner
1996). Small-scale velocity gradients have been inferred from methyl
cyanide observations (Olmi et al.\ 1996) but thus far the
interpretation is ambiguous. Subarcsecond observations of the (4,4)
inversion transition of ammonia by Cesaroni et al.\ (1998, hereafter
C98) represent the highest resolution molecular-line images published
to date, resolving the hot cores in G10.47+0.03, G29.96$-$0.02 and
G31.41+0.31. C98 also observed temperature and velocity gradients in
these three sources and concluded that the hot cores probably
encompass rotating flattened structures, perhaps disks.

Further evidence for the presence of embedded sources within hot cores
came from Testi et al.\ (2000) who discovered a faint 3.5-cm radio
continuum source within G9.62F, while Hofner, Wiesemeyer \& Henning
(2001) detected HCO$^+$ line wing emission with a spatially bipolar
distribution, interpreted as a molecular outflow, in the same hot
core. Beltr\'an et al.\ (2004) detected a velocity gradient in
G31.41+0.31 which they interpreted as rotation about a central source
driving a bipolar flow (seen in \thco : Olmi et al.\ 1996).

In order to shed further light on the nature of hot cores, we have
begun a high-resolution study with the Berkeley-Illinois-Maryland
Association (BIMA) millimeter array of a sample of four hot core
sources: G9.62+0.19, G10.47+0.03, G29.96$-$0.02 and G31.41+0.31
(hereafter G9.62, G10.47, G29.96 and G31.41 respectively). In this
paper we present observations of the H$_2$S 2(2,0)\too 2(1,1) and
\ceo\ 2\too 1 lines in order to explore the kinematics of the gas
associated with the hot cores. In a companion paper, Wyrowski, Gibb \&
Mundy (2004, hereafter Paper \wgm) present observations and modeling
of the continuum emission at 2.7 and 1.4 mm with sub-arcsecond
resolution of these objects. They find that in all cases but G10.47
the millimeter emission is clearly offset from the free-free emission
seen at centimeter wavelengths and that the strong 1.4-mm continuum
emission is evidence that these hot cores are massive, dense cores.

\hts\ is one such saturated species mentioned above, only forming on
dust grains or by hydrogenation reactions in hot shocked gas (e.g.\
Charnley 1997).  Models of the gas-phase chemistry involving \hts\
predict that over a period of $\sim$10$^4$ yr \hts\ reacts with
oxygen-bearing species to form the daughter products SO and SO$_2$,
causing the gas-phase abundance of \hts\ to decline from $\sim
10^{-7}$ to $10^{-10}$ or lower (e.g. Rodgers \& Charnley 2003;
Wakelam et al.\ 2004). Thus its distribution will trace the regions
where dust grains have been recently subject to sufficient heating to
evaporate material from the grain mantles (Hatchell, Roberts \& Millar
1999), either thermally by the radiation from nearby OB stars or
non-thermally through the action of shocks in an outflow. The gas
phase abundance of \hts\ is not well known, with a number of studies
giving values in the range $\sim 10^{-10}$ to $10^{-7}$ (van der Tak
et al.\ 2003; Minh et al.\ 1991). \hts\ has been observed in the
outflow associated with the Orion hot core by Minh et al.\ (1990) who
derive abundances as high as a few times $10^{-6}$ relative to
H$_2$. As we show below, we detect high-velocity \hts\ emission which
also originates in outflows from sources embedded within the hot
cores.

\section{Observations}

The BIMA array (Welch et al.\ 1996) was used in A (except G9.62), B, C
and D configurations on various dates between 1999 May and 2001
February. Nine antennas of the array were employed, tuned to a
lower-sideband frequency of 216.7 GHz (corresponding to a wavelength
of 1.4 mm). The intermediate frequency of 1.5 GHz places the upper
sideband at 219.7 GHz. The correlator setup included one spectral
window tuned to the frequency of the \hts\ 2(2,0)\too 2(1,1) at
216.7104 GHz with \ceo\ 2\too 1 at 219.5603 GHz falling in the upper
sideband. The observing bandwidth was 50 MHz, with a spectral
resolution of 781 kHz (corresponding to 68.7 and 1.07 \kms\
respectively).

Antenna gains were checked with observations of planets and found to
be accurate within 20 per cent. Phase calibration involved observation
of the quasars 1733$-$130, 1743$-$038 and 1751$+$096 for 3--4 minutes
every 15--30 minutes (depending on array configuration). Observations
made in the A configuration (as well as the B configuration for G9.62
and G29.96) also utilized fast switching between source and calibrator
every 1.5--2.5 minutes to track the phases (see Looney, Mundy, \&
Welch 2000 for further details). The continuum emission was
subtracted from the visibility data using channels free from line
emission. The continuum emission from each source at 1.4 mm is strong
enough to permit self-calibration. The upper and lower sidebands were
self-calibrated separately and the solutions applied to the respective
line data.

The data reduction was performed using the MIRIAD package (Sault et
al.\ 1995). A Briggs robust parameter of between 0 and 2 was used for
image construction and analysis. For the purposes of presenting the
data, the highest resolution with good signal-to-noise has been used
(with a robust value of typically 0.25--0.5). Table \ref{sources}
lists the relevant source details, including the resolution of the
images presented. Flux densities in Jy may be converted to brightness
temperatures in K assuming the following conversion factors for \ceo\
and \hts\ respectively -- 2.5 \& 3.4 (G9.62), 3.7 \& 7.3 (G10.47),
3.9 \& 4.3 (G29.96) and 6.4 \& 9.4 (G31.41).

\section{Results}

\begin{figure}
\figurenum{1}
\epsscale{0.9}
\plotone{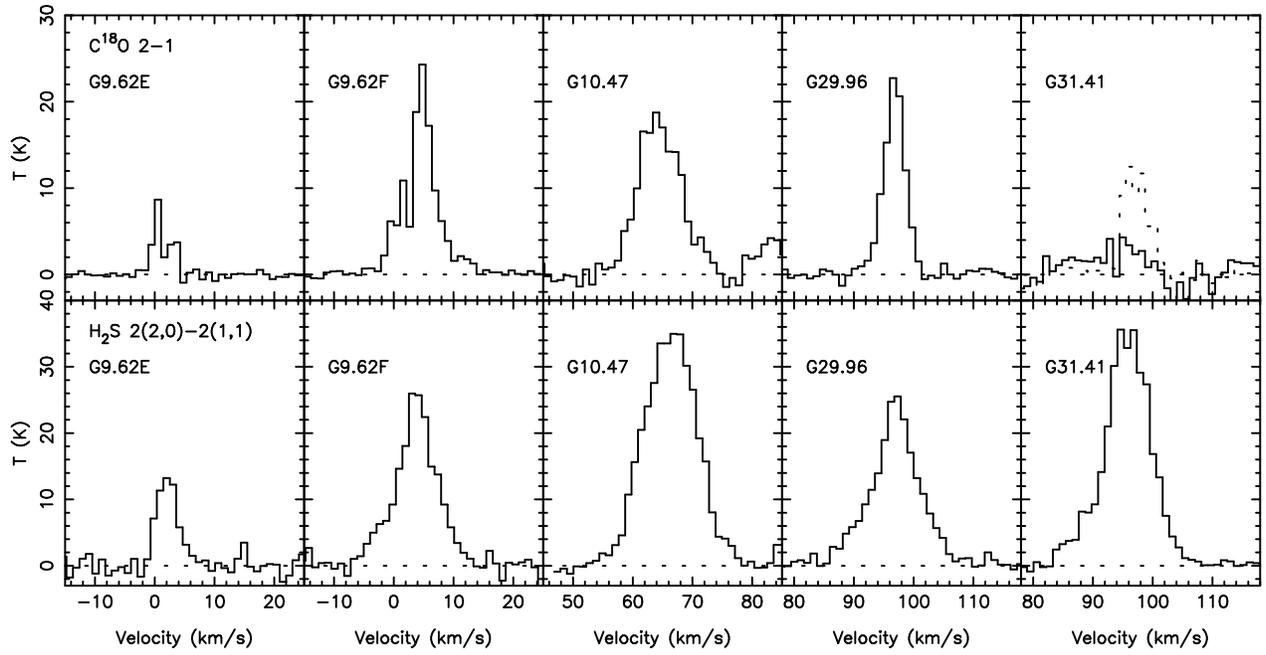}
\caption{\ceo\ and \hts\ Spectra toward the peak positions in each source. The
  vertical scale  for each transition is the same in each panel to show the relative
  strength of the line toward each source. The dotted \ceo\ profile for
  G31.41 represents the spectrum from the northern peak. }
\label{spectra}
\end{figure}

In this section we describe the main observational results, place them
in context with previous observations and present the analysis
alongside. \ceo\ and \hts\ spectra for each source are shown in Figure
\ref{spectra}, clearly showing the high-velocity wings in the \hts\
spectra toward three of the four target regions. The wings are most
prominent toward G9.62 and G29.96, while the broadest lines are seen
toward G10.47 and G31.41. Each source will now be described in more
detail.

\subsection{G9.62+0.19}

The region around G9.62+0.19 is known to have nearly a dozen \HII\
regions (Cesaroni et al.\ 1994; Testi et al.\ 2000), many of which are
compact or ultracompact. Components D--G appear to lie within an
elongated core, traced by CH$_3$CN and continuum emission at 2.7 and
1.4 mm (Hofner et al.\ 1996; Paper \wgm).  The brightest dust emission
at 1.4 mm coincides with the centimeter source G9.62F, with a
secondary maximum toward G9.62E. Testi et al.\ (2000) show that the
centimeter fluxes of G9.62F are consistent with an embedded B1--B1.5
star either exciting an ultracompact \HII\ region or driving an
ionized stellar wind.

\begin{figure}
\figurenum{2}
\epsscale{0.4}
\plotone{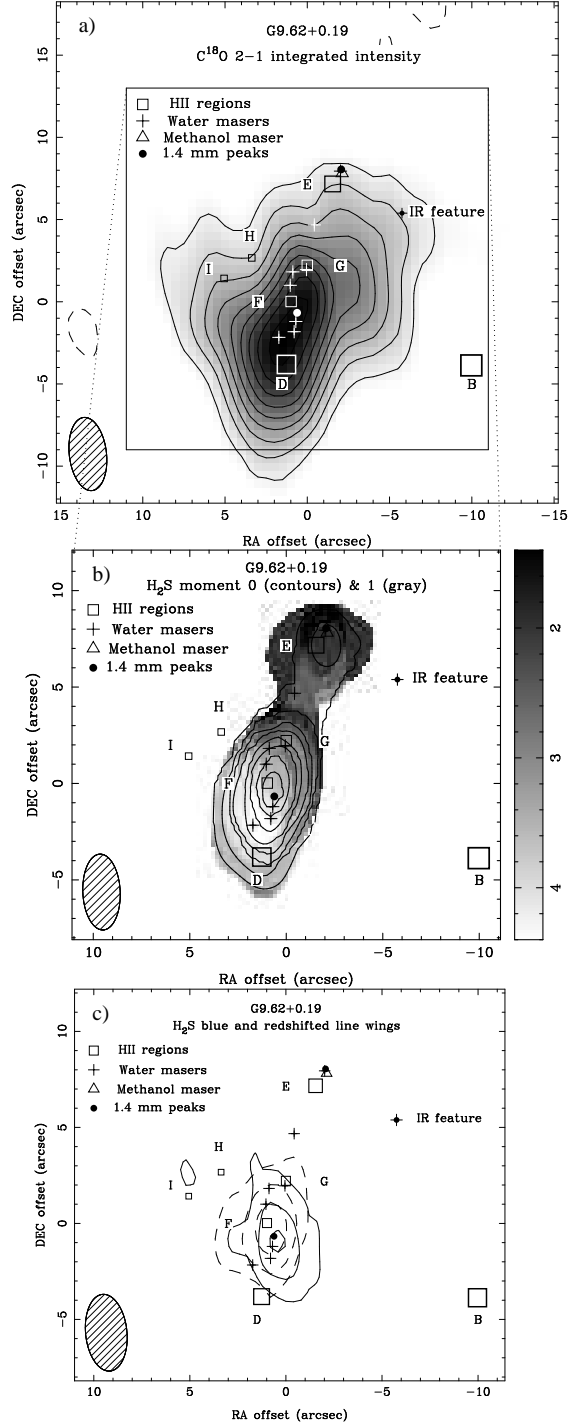}
\caption{\small a) \ceo\ 2\too 1 integrated intensity towards G9.62.
  \HII\ regions are marked with open squares and water masers from
  Hofner \& Churchwell (1996) are marked by plus signs. The IR feature
  of Persi et al.\ (2003) is marked by a plus sign with a small filled
  circle. The size of the open squares reflects the centimeter-wave
  flux density: a larger square represents a greater flux density.
b) Contours of the zeroth moment (clipped at 2.0 \jyb)
  superimposed on a color-scale representation of the first moment
  (clipped at 2.5 \jyb). Contours begin at 6
  Jy\,beam$^{-1}$\,km\,s$^{-1}$ and increase in steps of 4
  Jy\,beam$^{-1}$\,km\,s$^{-1}$. The grayscale scale varies linearly
  between 1.4 km\,s$^{-1}$ (black) and 4.2 km\,s$^{-1}$ (white).
The beam is marked by
the hatched ellipse in the bottom left-hand corner of the image. 
c) Superposition of blue-shifted  (solid contours) and
  red-shifted \hts\ emission (dashed contours). Average emission over
  a 4 \kms\ -wide range centered at $-$4 \kms\ (blue) and 10 \kms\ 
  (red).  Contours increase in steps of 0.36 Jy\,beam$^{-1}$ starting
  at 0.54 Jy\,beam$^{-1}$. }
\label{g962}
\end{figure}

\begin{figure}
\figurenum{3}
\epsscale{0.9}
\plotone{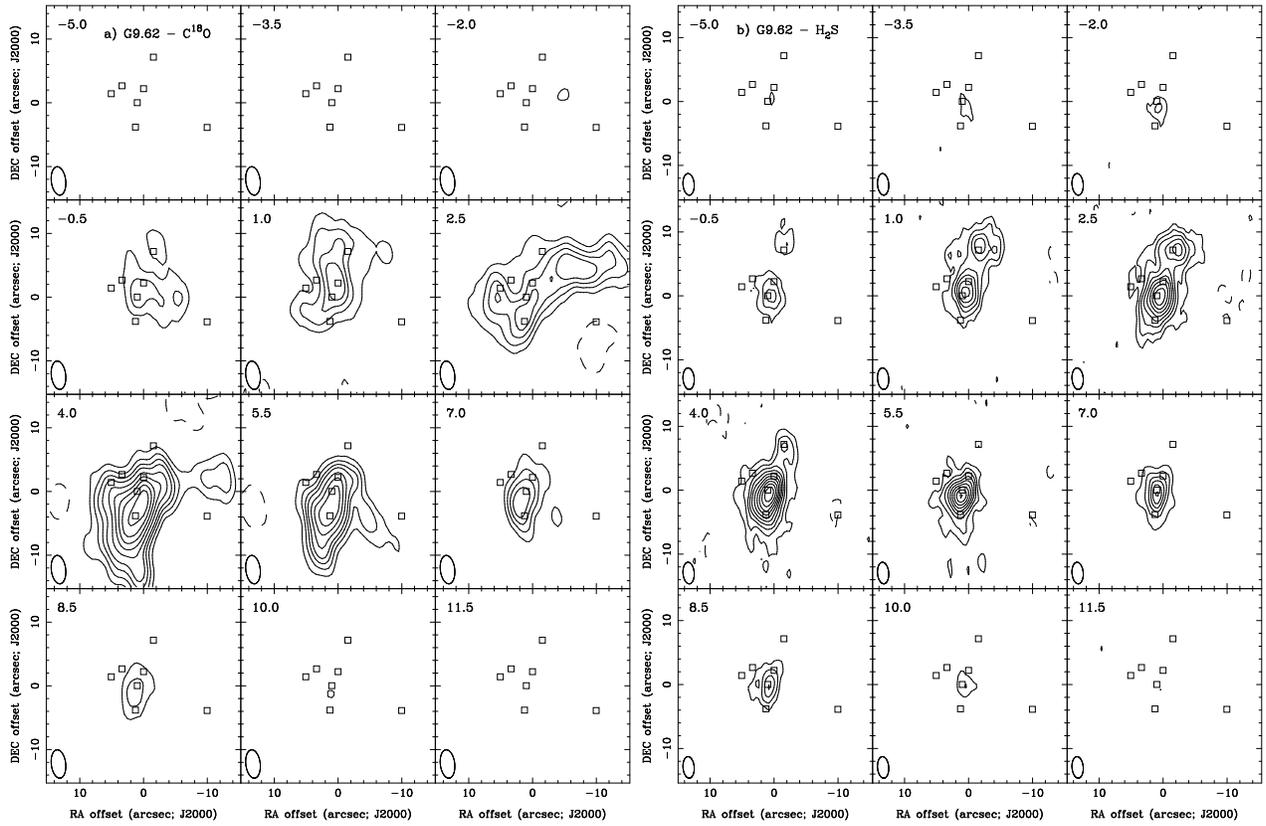}
\caption{G9.62 channel maps of a) \ceo\ and b) \hts\ emission from
  $-$5 to 11.5 \kms\ in steps of 1.5 \kms. Contours begin at 1.2/0.45
  \jyb\ and increase in steps of 0.8/0.30 \jyb\ for \ceo\ and \hts\
  respectively.  \HII\ regions are marked by open squares.
  The beam is marked by the ellipse in the bottom
  left-hand corner of the image. }
\label{g962chans}
\end{figure}

The \ceo\ 2\too 1 emission towards G9.62 defines a single elongated
core encompassing components F and D (Figure \ref{g962}a). Component E
lies towards the edge of the integrated emission but does not stand
out as a separate core. The emission shows a number of weaker
extensions which do not appear to be associated with any of the known
\HII\ regions. The overall distribution closely follows that of the
\ceo\ 1\too 0 emission shown by Hofner et al.\ (1996). Channel maps
(Figure \ref{g962chans}a) reveal what may be the limb of a shell of
gas defining the interface between the extended \HII\ region B and the
molecular cloud. A velocity gradient exists from north to south (blue
to red) as seen previously by Hofner et al.\ (1996) in their 1\too 0
data.

The \ceo\ spectra show self-absorption dips toward the center of the
core. Comparing the total flux with the single-dish measurements of
Hatchell et al.\ (1998b) show that our observations recover all of the
\ceo\ emission to within the calibration uncertainties. Therefore, it
is likely that these dips are genuine and not a result of missing flux
in the map. The presence of the self-absorption makes it difficult to
reliably estimate linewidths, but away from the core center the
typical linewidth is $\sim$4.0 \kms. There appears to be a weak red
wing on the \ceo\ spectrum toward component F, as also observed by
Hofner et al.\ (1996). However, a map of the blue and redshifted
emission does not show clear evidence for an outflow.

The total integrated \hts\ emission in G9.62 also defines an elongated
core (see Figure \ref{g962}b), but it is clear that the denser gas
traced by \hts\ is concentrated toward components E and F with two
clear maxima seen toward these sources. The emission also extends to
cover components D and G but it is not possible to distinguish
separate cores. The \hts\ emission peaks coincide with the 1.4-mm
continuum sources presented in Paper \wgm. No \hts\ emission is
detected toward components B, C, H or I. No emission was detected
toward the extended near-infrared feature discovered by Persi et al.\
(2003).

The \hts\ lines tend to be broader than the \ceo\ lines, indicating
their formation in more turbulent regions closer to the center of the
core, and are broadest at the position of component F with a
full-width-at-half-maximum (FWHM) of 6.8 \kms. The FWHM linewidth at E
is 3.1 \kms, 5.2 \kms\ at G and 3.8 \kms\ toward component D. The
spectrum toward F shows both blue and redshifted wing emission, and a
map of the blue-shifted emission between $-$6 and 0 \kms\ reveals a
well-defined lobe of emission slightly displaced from the center of
core F. The redshifted gas (integrated between 8 and 13 \kms) defines
a counterpart to the blue lobe, and is centered on component F: see
Figure \ref{g962}c. Most of the H$_2$O masers near G9.62F are
redshifted, suggesting that they arise in the outflow. The outflow
lobes do not show the same position angle nor do they peak at the same
position as those seen in the HCO$^+$ data of Hofner et al.\ (2001)
but they are in the same sense: blueshifted gas to the SW, redshifted
to the NE. The \hts\ results therefore reinforce the interpretation
that F is the hot core source and that it contains a young high-mass
protostellar object.

Like the \ceo , the \hts\ data also delineate a clear velocity
gradient along the major axis of the core, also seen previously by
Hofner et al.\ 1996), shown in a grayscale representation in Figure
\ref{g962}b (although not so obvious from the channel maps in Figure
\ref{g962chans}b). The \hts\ outflow emerges at an angle to the long
axis of the core. The position angle of the outflow is 10.5 degrees
east of north; that of the core is $-$13 degrees. (The position angle
of the HCO$^+$ outflow is 28 degrees.) The orientation of the blue and
red lobes of the outflow defines a velocity gradient which is in the
{\em opposite} sense to that seen along the major axis of the core,
i.e.\ redshifted in the south and blueshifted to the north and west,
further supporting our interpretation that the line wings arise in a
bipolar outflow driven by an embedded source within F.
Hofner et al.\ (2001) present arguments for the outflow being viewed
pole-on. In this geometry it is unlikely that significant rotation can
be detected. An alternative (and perhaps more likely) possibility is
that the velocity gradient is due to the superposition of three cores
(housing components D, F and G) which lie at different
velocities. Higher angular resolution would enable this hypothesis to
be tested.

While G9.62E is clearly detected in \hts\ emission, no line wings are
observed which implies that either it has yet to form an outflow or it
has evolved beyond the outflow stage. Given its well-defined \HII\
region (stronger than F) and weaker dust emission than component F, it
is likely that G9.62E is more evolved than G9.62F and has begun to
clear away its parent core. 

\subsection{G10.47+0.03}

G10.47+0.03 houses a group of four ultracompact \HII\ regions, three
of which (A, B1 and B2) lie within a region 2 arcsec across (C98). C98
argue that the majority of the NH$_3$ (4,4) emission originates behind
the three \HII\ regions, with component A being the least
embedded. The ammonia peaks approximately 0.5 arcsec east of B1. The
fourth \HII\ region, G10.47C, lies outside the area of ammonia
emission. In this interpretation, the radio continuum sources are not
deeply embedded within the hot core, instead lying close to the front
surface.

\begin{figure}
\figurenum{4}
\epsscale{0.4}
\plotone{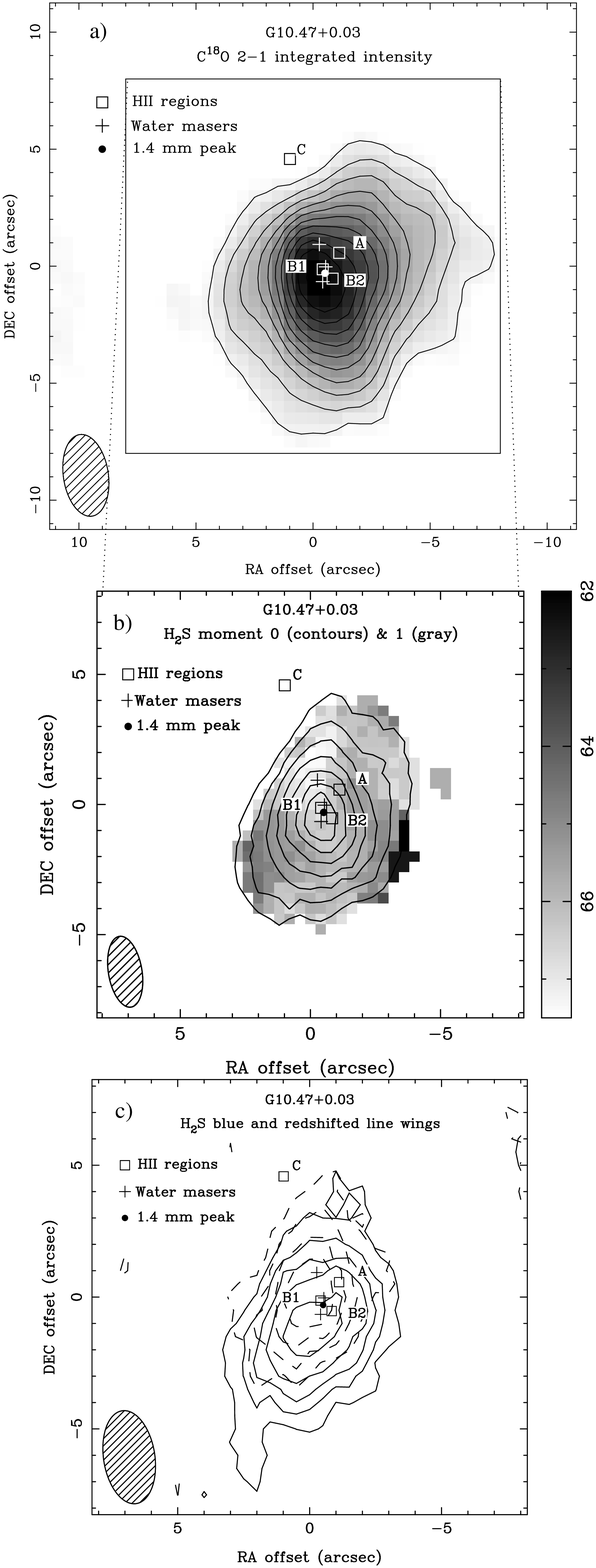}
\caption{a) G10.47 moment zero contours (clipped at a flux density of
  1.0 \jyb) superimposed on a grayscale of the first moment (clipped
  at 2.0 \jyb\ to isolate the line core). Contours begin at 10
  \jyb\,\kms\ increasing in steps of 10 \jyb\,\kms. The peak is 78
  \jyb\,\kms. Ultracompact \HII\ regions are marked by open squares
  and water masers by plus signs.  The beam is marked by the ellipse
  in the bottom left-hand corner of the image.  b) G10.47 high
  velocity emission averaged over the velocity ranges 74 to 76 \kms\
  (red) and 58 to 60 \kms\ (blue). Redshifted gas is shown by the
  dashed contours and blueshifted gas by the solid contours.  Contour
  levels are the same for each lobe, beginning at 0.54 \jyb\ in
  increasing in steps of 0.36 \jyb.}
\label{g1047}
\end{figure}

\begin{figure}
\figurenum{5}
\epsscale{0.9}
\plotone{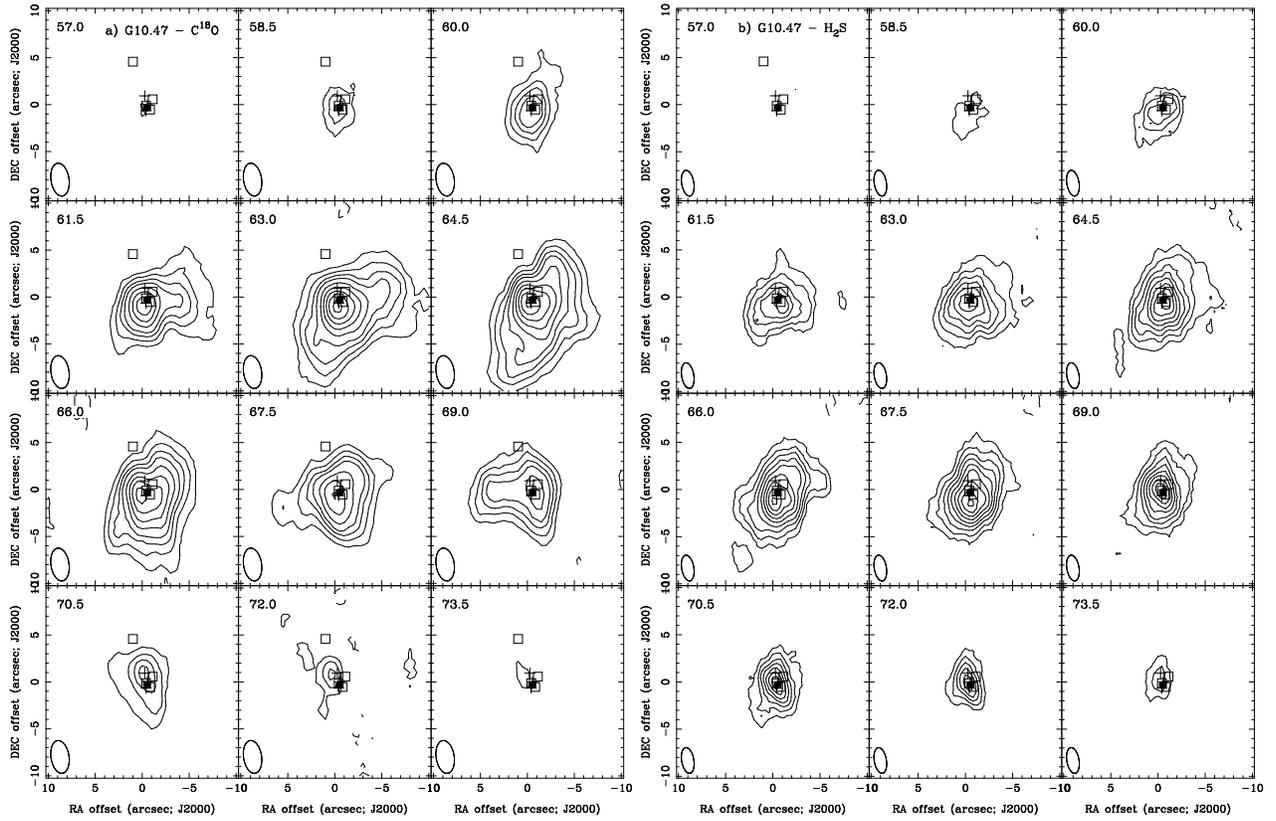}
\caption{G10.47 channel maps of a) \ceo\ and b) \hts\ emission between
  57 and 73.5 \kms. Contours begin at 0.6/0.75 \jyb\ increasing in
  steps of 0.4/0.5 \jyb\ for \ceo\ and \hts\ respectively.  Symbols
  are as in Figure \ref{g1047}. The beam is marked by the ellipse in
  the bottom left-hand corner of the image.}
\label{g1047chans}
\end{figure}

The \ceo\ emission (Figure \ref{g1047}a) defines a single compact core
peaking on the \HII\ regions B1 and B2, with an extension to the
west. No emission is detected toward component C. A number of features
including the westward extension show up more clearly in the channel
maps (Figure \ref{g1047chans}a). The westward extension covers the
velocity range 61.5 to 64.5 \kms\ (curving further north at 64.5
\kms), while between 64.5 and 69 \kms\ a corresponding feature exists
in the south-east to east, with a well-defined tongue of emission
extending to the east at 69 \kms. Such features could represent a
bipolar outflow, or a rotating (fragmented) core. We rule out outflow
since the spectra to the east and west of center peak at different
velocities, and thus the east and west extensions are not due to
high-velocity line wings. Therefore it seems likely that these
features define a gradient in the bulk of the core, and may represent
smaller and less-massive individual cores. 

The \ceo\ spectra are single peaked across the entire region, and
broadest at the position of the hot core. The linewidths at the
positions of the east and west components are much narrower than
toward the hot core ($\sim$3 \kms). Comparing our data with those of
Hatchell et al.\ (1998b) again shows that we detect all of the \ceo\
flux to within the calibration uncertainties.

The \hts\ line is only detected in emission towards G10.47 and we do
not see the same bow-shape seen in the ammonia of C98 caused by
absorption against the \HII\ regions. The integrated \hts\ emission
(Figure \ref{g1047}b) shows a single-peaked core elongated in a NW-SE
direction at the lowest contours, but twists to a NE-SW direction in
the highest contours, encompassing the UC\HII\ regions, B1 and
B2. This suggests that there are two separate cores containing B1 and
B2 which we are unable to resolve. Like C98, we also do not detect any
molecular line emission toward G10.47C.

The brightest emission lies slightly south of B1/2, and extends
further south by $\sim$2 arcsec, consistent with elongation in dust
continuum (Paper \wgm). The spectra toward each of the \HII\ regions
are flat-topped which could be due to either the presence of two
velocity components which we are unable to resolve (perhaps from an
expanding shell) or high optical depths in the \hts\ line. In our
discussion below (Section \ref{columns}) we conclude that the \hts\ is
most likely to be optically thick.

The \hts\ covers a velocity range from 58 to 74 \kms\, a similar range
to the \ceo. Unlike the other sources in this sample, the \hts\ line
does not show the same clear high-velocity wings (Figure
\ref{spectra}). However, the \hts\ lines toward G10.47 are the
broadest in our survey with a peak FWHM of 10.5 \kms. From channel
maps it is possible to distinguish two velocity components, although
these do not correspond to any of the kinematic features of the \ceo\
data. The first component extends in the NE--SW direction and covers
the range 61 \kms\ to 73 \kms. A second component extends NW--SE which
traces material between 59 and 70 \kms.  The NE-SW velocity component
is also seen in NH$_3$ (4,4) of C98. A secondary peak is observed at
(3,$-$7) arcsec offset over the velocity range 65 to 67 \kms,
coincident with one of the \ceo\ cores desrcibed above.

While the \hts\ spectra do not show clear high-velocity line wings,
the extreme red and blue shifted emission does show a slight
north-south separation, although it is less than one beam diameter
(Figure \ref{g1047}c). The center of this `outflow' lies $\sim 1''$
north of the the 1.4-mm continuum peak and the UC\HII\ regions B1 and
B2. The fact that all the water masers are aligned on a north-south
axis lends support to an outflow interpretation. The outflow
interpretation gains further support from the fact that Olmi et al.\
(1996) detected a velocity gradient in \thco\ consistent with a
north-south outflow (in the same sense as we observe in \hts), as well
as an east-west velocity gradient in CH$_3$CN, i.e.\ perpendicular to
the proposed outflow (although see \S\ \ref{ch3cn} below). The small
offset between the lobes means that we have not resolved a flow
direction, or that there is no outflow and we are be seeing the front
and back surfaces of an expanding shell of gas centered on the \HII\
regions. We therefore conclude that there is only weak evidence for
either outflow or rotation toward G10.47.

\subsection{G29.96$-$0.02}

The ultracompact \HII\ region in G29.96$-$0.02 shows a classic
cometary shape (Wood \& Churchwell 1989), which is also evident in our
2.7 and 1.4-mm continuum data (Paper \wgm). C98 also observed G29.96
in ammonia (4,4) and detected a single peak coincident with a group of
water masers, but offset from the \HII\ region. Dust emission peaks at
the position of the ammonia and water masers (Olmi et al.\ 2003; Paper
\wgm) confirming the location of the hot core.

\begin{figure}
\figurenum{6}
\epsscale{0.4}
\plotone{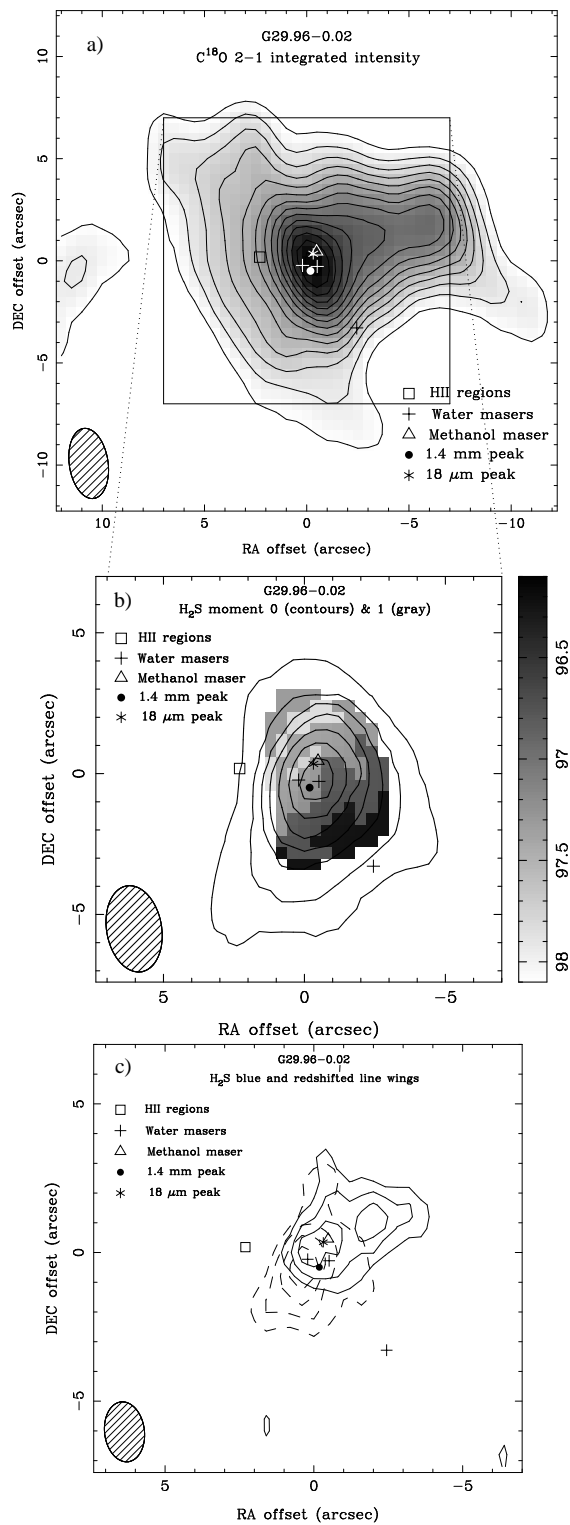}
\caption{a) Contours of zeroth moment superimposed on the first moment
  grayscale image (clipped at a flux density of 2 Jy\,beam$^{-1}$ to
  isolate the line core). Note the SW--NE gradient in the first moment
  data perpendicular to the direction of the outflow lobes, suggestive
  of a rotating core. b) Red-shifted (dashed contours) and
  blue-shifted (solid contours) H$_2$S emission averaged over a 4-\kms
  -wide interval centered at 90 \kms\ (blue) and 106 \kms\ (red).  The
  18-$\mu$m peak is almost coincident with the methanol maser (De
  Buizer et al.\ 2002) which lies within the blue lobe of the
  outflow. Contours for each lobe begin at 0.1 \jyb\ and increase in
  steps of 0.1 \jyb. The beam is marked by the ellipse in the bottom
  left-hand corner of the image.}
\label{g2996}
\end{figure}

\begin{figure}
\figurenum{7}
\epsscale{0.9}
\plotone{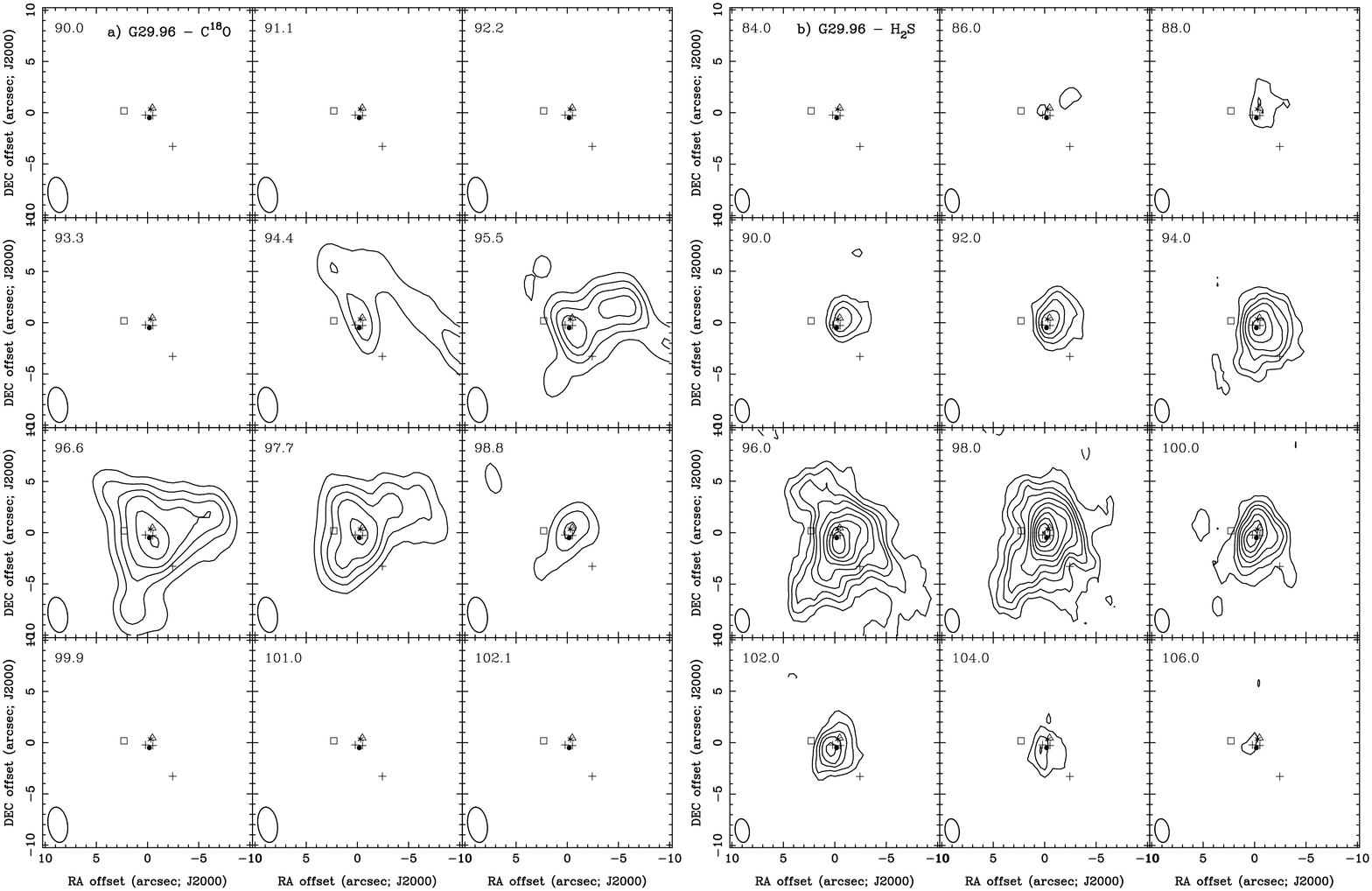}
\caption{G29.96 channel maps of \hts\ emission from 90 to 102.1 \kms\
  (\ceo) and 84 to 106 \kms\ (\hts) in steps of 1.1/2 \kms (\ceo/\hts\
  resepctively). Contours begin at 1.2/0.3 \jyb\ and increase in steps
  of 0.8/0.2 \jyb\ for \ceo\ and \hts\ respecetively. The 1.4-mm
  continuum peak is marked by a filled circle, the position of the
  UC\HII\ region is marked by an open square, water masers are marked
  by plus signs, the methanol maser group of Minier et al.\ (2002) are
  marked by an open triangle. The beam is marked by the ellipse in the
  bottom left-hand corner of the image.}
\label{g2996chans}
\end{figure}

The \ceo\ emission toward G29.96 defines an extended core with
multiple peaks (Figure \ref{g2996}a). There is general good agreement
with the 1\too 0 data of Olmi et al.\ (2003), although they do not
detect the westward ridge we see in our data. The \ceo\ emission shows
a well-defined edge extending from north-east to south-west, which is
likely associated with the boundary between the molecular cloud and
the ultracompact \HII\ region. This is seen more clearly in the
channel maps, Figure \ref{g2996chans}a. This fact indicates that the
hot core and the ultracompact \HII\ region are embedded within the
same molecular cloud. The lone water maser $\sim$4.5$''$ to the SW
does not appear to be associated with any other peaks in either the
\ceo\ or \hts\ emission.

The \ceo\ emission peaks on the hot core, decreasing in intensity to
the west, where at least two further cores can be discerned. These
cores have not been resolved by any previous observations. There is a
very slight velocity gradient along this ridge with the more distant
core being marginally blue-shifted relative to the hot core (spectra
peak at 96.8, 96.5 and 96.1 \kms\ from east to west).

The \ceo\ lines are relatively narrow across much of the source,
peaking at the hot core position. G29.96 has the smallest range of
\ceo\ emission of our sample of sources, covering only 5 \kms (Figure
\ref{g2996chans}a). The linewidths decrease along the westward ridge
away from the hot core (4.0, 3.5 and 3.2 \kms\ for the hot core, W1
and W2 respectively). Once again, there is good agreement between our
\ceo\ flux and the single-dish measurement of Hatchell et al.\ (1998b).

The integrated \hts\ emission shows a single maximum coincident with
1.4-mm continuum peak (Figure \ref{g2996}b).  Channel maps show that
the \hts\ is extended over a region $\sim 15''$ in diameter and
appears to wrap around the head of the UCHII region (Figure
\ref{g2996chans}b), in a similar manner to the \ceo.  The \hts\
emission in G29.96 is the most extensive of all our four sources, and
shows extensions to the west and south-west which may be further
individual cores.

The \hts\ spectra (Figure \ref{spectra}) show strong line wings
extending to $\pm$10 \kms\ across a wide area. Redshifted wings are
seen in the south-east while blue-shifted wings extend to the
north-west. Figure \ref{g2996}c shows the integrated red and
blueshifted emission and clearly shows the spatial bipolarity of an
outflow.  The outflow center appears to lie closer to the ammonia peak
of C98 than our 1.4-mm peak. This may indicate that there are multiple
sources within the G29.96 hot core: higher resolution observations are
clearly needed. The orientation of the blue lobe is along line of
methanol masers seen by Minier, Booth \& Conway (2002). The methanol
masers are also blueshifted, increasingly blueshifted with distance
from the outflow center. 

Figure \ref{g2996}b also shows an image of the first moment of the
\hts\ emission (clipped at 
a level which excludes the wing emission), revealing a velocity
gradient along a SW to NE axis, blue-shifted in the SW and redshifted
toward the NE. There is evidence of this same gradient in the \ceo\
data as well. Figure \ref{g2996chans}a shows that the \ceo\ peaks to
the SW of the hot core at 95.5 \kms\ and to the NE at 97.7 \kms. It is
tempting to interpret this as rotation of the hot core, especially
since the gradient is observed to be perpendicular to the outflow
direction. However we regard this conclusion as somewhat tentative and
clearly requires further observations of high-density tracers which
are not contaminated by outflow. Under the assumption that we are
detecting rotation the velocity gradient is $\sim$11 \kms \,
pc$^{-1}$, corresponding to a rotation period of $8.9\times 10^4$ yr.

\subsection{G31.41+0.31}

The region associated with G31.41+0.31 houses an ultracompact \HII\
region along with a diffuse shell of extended radio continuum emission
(Wood \& Churchwell 1989). Hofner \& Churchwell (1996) detect a group
of 8 water masers, none of which coincide with the UC\HII\
region. However, the ammonia observations of C98 reveal a compact core
of molecular gas clearly coincident with the water masers, but with no
associated centimeter continuum emission.

\begin{figure} 
\figurenum{8}
\epsscale{0.4}
\plotone{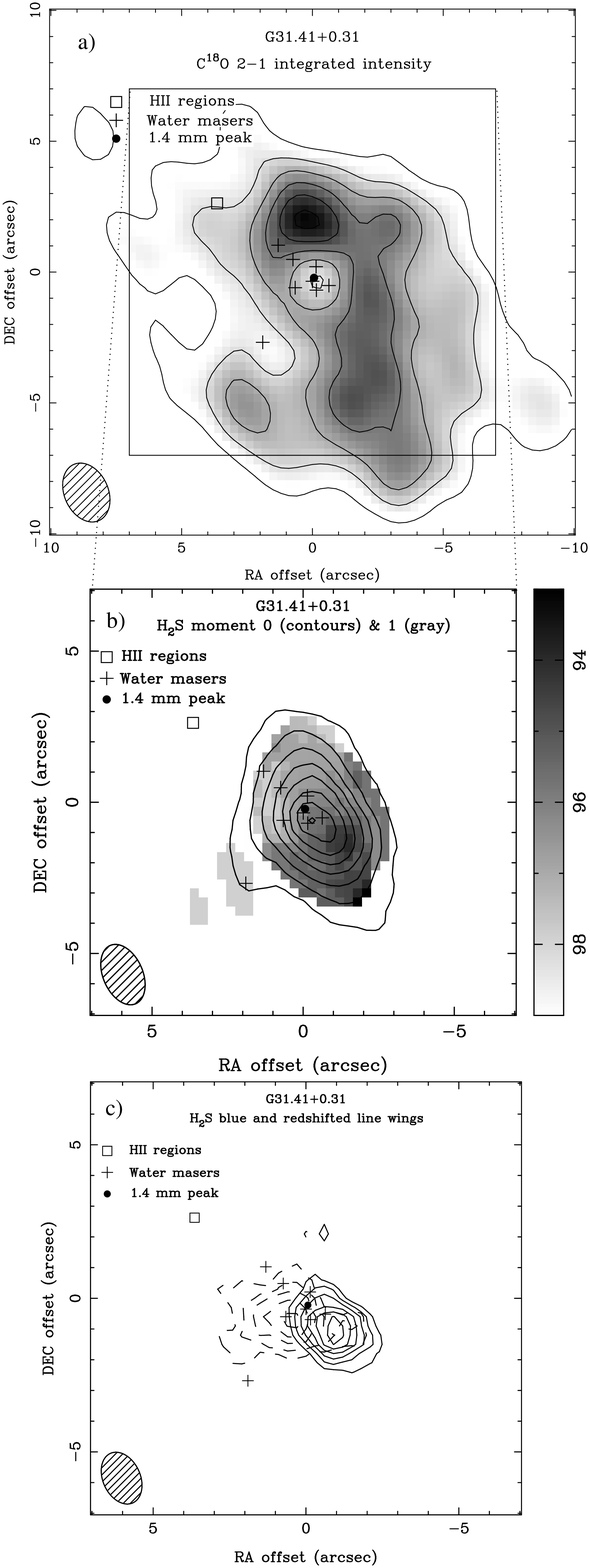}
\caption{a) Moment zero (integrated intensity, clipped at 0.5 \jyb)
  contours superimposed on grayscale of the first moment image
  (clipped at 1.5 \jyb). Contours start at 4 \jyb\,\kms, increasing in
  steps of 5 \jyb \, \kms. The peak is 41 \jyb\,\kms. Grayscale is
  linear from 92.5 (blue) to 98.9 \kms\ (red) as shown on the sidebar.
  b) G31.41 redshifted contours (red) and blueshifted contours (blue)
  averaged over a 4-\kms -wide interval centered at 89 \kms\ (blue)
  and 105 \kms\ (red). For each lobe, the contours begin at 0.2 \jyb\
  and increase in steps of 0.1 \jyb. }
\label{g3141}
\end{figure}

Our \ceo\ emission shown in Figure \ref{g3141}a defines an elongated
core, mostly oriented north-south but with a clear contribution from a
diffuse, more extended component oriented NE-SW. Most notable is the
fact that the \ceo\ does not peak at the hot core position; in fact
the hot core is seen as a local minimum in the \ceo\ emission. A
comparison of our data and the JCMT measurement of Hatchell et al.\
(1998b) shows that we only detect of order 70\% of the \ceo\ flux. It
may be that the \ceo\ is not strongly peaked on the hot core position
and is thus filtered out by our interferometric
observations. Alternatively, the brightness temperature of the
continuum emission may be sufficiently high (at 27.5 K for the
naturally-weighted data) that the \ceo\ is simply not detectable above
this level. There are a number of peaks evident in the integrated map,
although it is not certain if they represent independent cores given
that some of the flux is missing from the image.

Channel maps reveal the structure of the core. Figure \ref{g3141}a
shows the presence of a clear velocity gradient along the major axis
of the \ceo\ core, blue-shifted in the south and red-shifted to the
north. The individual peaks seen in the integrated intensity map are
clearly seen in the channel maps.

Figure \ref{g3141}b shows that the integrated \hts\ emission is
centered on the 1.4-mm and ammonia emission, with the water masers
mostly lying along an axis parallel with the observed extension of the
\hts\ core. The first moment image also shown in Figure \ref{g3141}b
shows a weak velocity gradient along the core (in the same sense as
the larger-scale \ceo\ gradient), which is tempting to interpret as
rotation. The high-velocity wing emission (shown in Figure
\ref{g3141}c) is predominantly east-west in orientation.  A
position-velocity cut along the core shows the velocity gradient more
clearly (see \S\ \ref{dynamics} below). We need further observations
of complementary tracers to isolate one velocity component from the
other. We note that the lower-resolution HCO$^+$ 1\too 0 and SiO 2\too
1 observations of Maxia et al.\ (2001) show the same distribution of
red- and blueshifted high-velocity gas, supporting the outflow
interpretation. The water masers appear to be associated with the
outflow since many of the maser spots are red-shifted.

\begin{figure}
\figurenum{9}
\epsscale{0.9}
\plotone{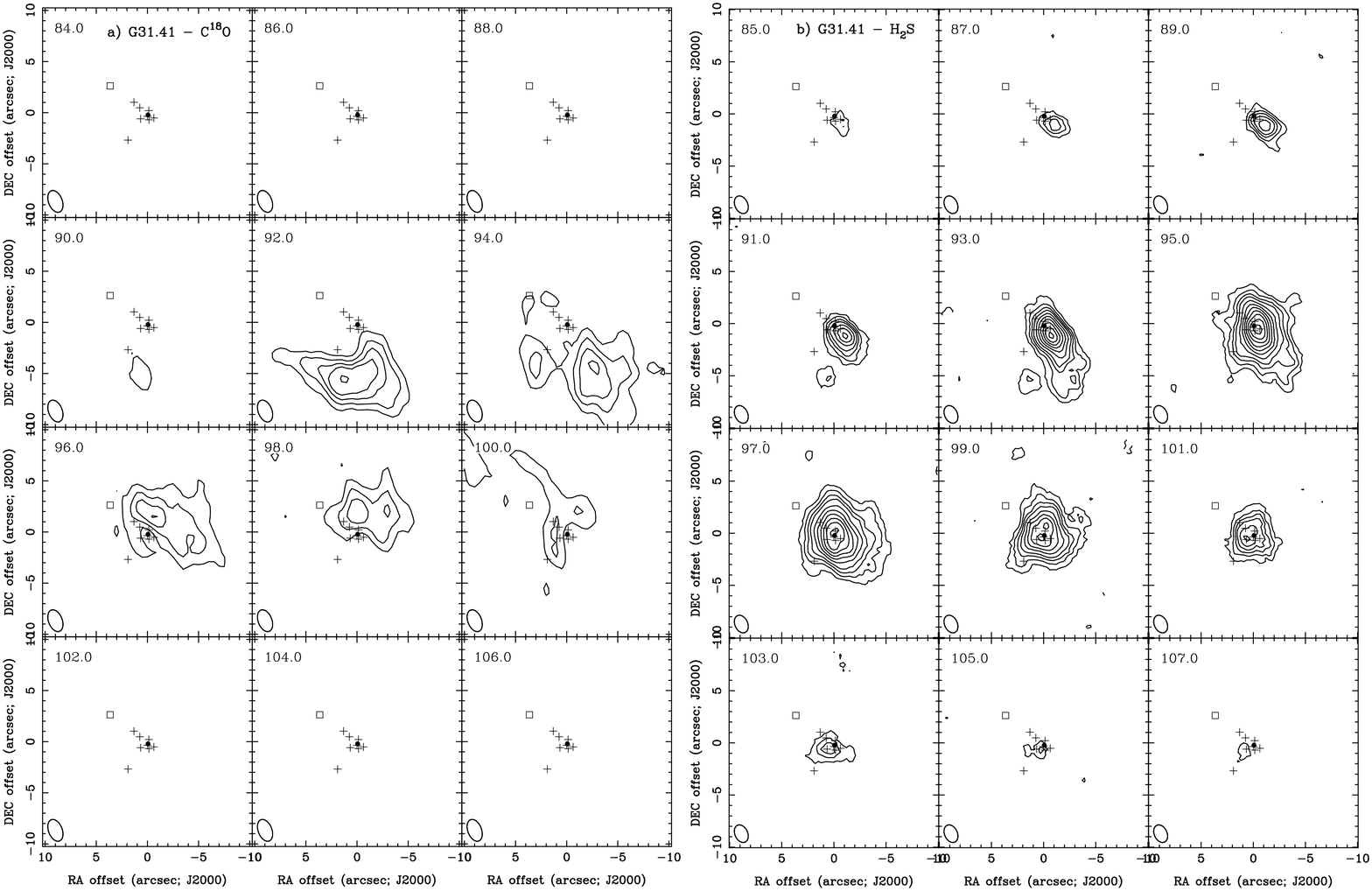}
\caption{G31.41 channel maps of a) \ceo\ and b) \hts\ emission from 85
  to 107 \kms\ increasing in steps of 2 \kms. Contours begin at
  0.45/0.3 \jyb\ and increase in steps of 0.3/0.2 \jyb\ for \ceo\ and
  \hts\ respectively.  Plus signs mark the location of water masers
  from Hofner \& Churchwell (1996), the position of the UC\HII\ region
  is marked by an open square and the filled circle marks the position
  of the 1.4-mm continuum peak. The beam is marked by the ellipse in
  the bottom left-hand corner of the image.}
\label{g3141chans}
\end{figure}

Channel maps (shown in Figure \ref{g3141chans}b) show that the \hts\
emission is largely extended in the same NE-SW direction at all
blue-shifted velocities. At the south end of the core, two secondary
peaks show up at 91 to 93 \kms, and it may be these components which
bias the first moment image toward greater blue-shifted velocities in
this regions.  These may represent separate independent cores since
our \ceo\ data show an extended maximum toward these positions (and at
the same velocity), as do the HCO$^+$ data of Maxia et al.\
(2001). The \hts\ spectra show varied line shapes across the core,
with clear blue- and red-shifted wing emission toward some positions,
and double-peaked line profiles toward the center of the
core. However, as described above for G10.47, the double-peaked
profiles may be due to self-absorption of an optically thick line.

\section{Analysis: Hot core masses and outflow parameters}

In this section we describe our analysis of the data in order to
derive various physical parameters of the hot cores. We have made
two-dimensional gaussian fits to the integrated-intensity images to
derive source dimensions and intensities. In most cases a gaussian fit
was a good representation of the observed emission. The deconvolved
source dimensions were estimated assuming a gaussian geometry: these
are the values listed in Table \ref{properties}. The mean linewidth
was derived from gaussian fits to spectra toward each source (also
given in Table \ref{properties}). Next, virial masses were calculated
assuming that each core can be approximated to a sphere of constant
density. If the density falls off with radius then, for typical
power-law indices, the virial masses will be smaller by up to 40 per
cent (see e.g.\ MacLaren, Richardson \& Wolfendale 1988). H$_2$ column
and volume densities were calculated from these virial masses, again
assuming constant density spheres. For \ceo\ we also calculated a
total H$_2$ mass based on the integrated intensity since the abundance
is assumed to be well known (2$\times$10$^{-7}$). These derived values
are listed in Tables \ref{cparameters} and \ref{hparameters}.

\subsection{Masses and column densities}
\label{columns}

There is generally good agreement between the virial mass estimates
from the \hts\ data and those derived from the \ceo. In the \ceo\
data, G29.96 has the lowest virial mass and G10.47 the largest. G10.47
also has the largest \hts -derived virial mass, although now G9.62E
has the lowest value. With the exception of G29.96, the \ceo\ virial
masses are larger than the \hts\ masses, a result of the larger core
size measured in the \ceo\ data. As expected, the \hts\ is tracing
denser gas than the \ceo, and yields larger estimates of the H$_2$
column density on account of the smaller source size.

The molecular hydrogen column densities inferred from the \hts\ virial
masses are exceedingly high. In particular, G10.47 has the highest
column density in our sample at $6\times 10^{24}$\,cm$^{-2}$,
corresponding to a visual extinction of 12\,600 magnitudes. Even at
mid-infrared wavelengths, the extinction remains significant. Using
the values given by Mathis (1990), this H$_2$ column density
corresponds to an extinction of 270--300 magnitudes at 20 $\mu$m. The
extinction extrapolated to 1.4 mm is only 0.17--0.20 magnitudes, or
essentially optically thin ($\tau \sim$0.25).  However, since the
grain properties in the dense regions toward these hot cores can be
different to those in the interstellar medium, the actual
submillimeter optical depth could be higher. The \hts\ virial masses
imply a mean density in the four hot core sources of order
$10^7$\,cm$^{-3}$, in good agreement with values derived by Cesaroni
et al.\ (1994).

We have calculated the \ceo\ and \hts\ column density toward the hot
core positions in each source (plus component E in G9.62), shown in
Tables \ref{cparameters} and \ref{hparameters}. We have assumed the
emission to be optically thin, and that the level populations can be
characterized by a single excitation temperature. The choice of
excitation temperature is not constrained since we only have a single
transition. For \ceo\ we have assumed values between 16 (equal to the
energy of the $J$=2 level to obtain the minimum value: Macdonald et
al.\ 1996) and 125 K (a reasonable upper limit based on ammonia and
CH$_3$CN estimates, e.g. Cesaroni et al.\ 1994; Olmi et al.\
1996). Using the H$_2$ column densities derived from the \ceo\ virial
masses to estimate an abundance of \ceo\ yields values smaller than
the canonical value of $2\times 10^{-7}$ for G10.47 and G31.41, but in
reasonable agreement for G9.62 and G29.96. Therefore, either the \ceo\
towards G10.47 and G31.41 is not optically thin, or we are resolving
out a significant portion of the \ceo\ emission. Since the agreement
with previous single-dish observations is good for G10.47, we conclude
that the \ceo\ is optically thick in this source. The C$^{17}$O and
\ceo\ data of Hofner et al.\ (2000) show that the \ceo\ 2\too 1 line
has moderate optical depth ($\sim$1--2) in an 11\arcsec\ beam for all
our sources. However, it is likely that the missing flux contributes
to the poor agreement for G31.41. We have also calculated masses from
the integrated intensity maps, assuming the standard \ceo\ abundance
and optically thin emission. Table \ref{cparameters} shows that there
is good agreement between the masses calculated in this way and the
virial masses for G9.62 and G29.96. The integrated masses are much
lower than the virial masses for G10.47 and G31.41, which is probably
due to the effects discussed above.

The observed \hts\ line brightnesses yield minimum excitation
temperatures of 18 to 40 K (Figure \ref{spectra}).  However, a strict
lower limit for \hts\ can be obtained assuming that $T_{\rm ex} =
2E_{\rm u}/3k = 56$\,K (Macdonald et al.\ 1996) where $E_{\rm u}$ is
the energy of the upper level, equivalent to 84 K in temperature units
for the \hts\ 2(2,0)\too 2(1,1) transition.  However, the exact value
is relatively unimportant as the derived \hts\ column density changes
by only 50 per cent over the temperature range 30 to 250 K. Therefore,
our choice of excitation temperature results in only a small
uncertainty in our calculations, and lower than the uncertainty due to
unknown optical depth. As a result, we only list a minimum value in
Table \ref{hparameters}. Examining the line intensities we see that
for an excitation temperature of 56 K, the lines in G10.47 and G31.41
are not optically thin ($\tau \sim 1$, and higher for lower excitation
temperature).

The \hts\ column density varies from $3\times 10^{15}$\,cm$^{-2}$
toward G9.62E to $4.1\times 10^{16}$\,cm$^{-2}$ toward G10.47. If we
assume that the virial masses calculated above are accurate
representations of the mass of the region containing the \hts\ then we
can estimate the abundance of \hts\ in each source. This yields
column-averaged \hts\ abundances of $1.2$--$6.8\times 10^{-9}$
relative to H$_2$. Bearing in mind the fact that the virial masses are
probably upper limits, and the fact that the \hts\ column density is a
lower limit, these abundances should themselves be regarded as lower
limits, although they are in good agreement with estimates from
Hatchell et al.\ (1998b) and van der Tak et al.\ (2003).

\subsection{Comparison with previous estimates}
\label{masses}

The first comparison we make is with our own BIMA continuum data
presented in Paper \wgm. To estimate the dust masses we assumed an
dust emissivity of 0.018 cm$^2$\,g$^{-1}$ at 1.4 mm (Ossenkopf \&
Henning 1994) and a temperature between 50 and 250 K. This yields dust
masses in the range 17 to 94 M$_\odot$ (G9.62F), 151 to 824 M$_\odot$
(G10.47), 36 to 198 M$_\odot$ (G29.96) and 156 to 850 M$_\odot$
(G31.41) with the larger values corresponding to the lower
temperatures. In all cases, the \ceo\ and \hts\ are considerably more
extended than the dust emission, with FWHM dimensions typically twice
as large as those of the dust cores.

One the whole, there is reasonable agreement between the dust masses
and the virial masses for both \ceo\ and \hts, at least for the lower
dust temperatures. If the dust is indeed as hot as 250 K, however,
then the discpreancy between the dust and molecular line estimates can
be large, ranging from a factor of 3 to more than an order of
magnitude difference. If the dust emission is optically thick, we will
underestimate the dust masses. Earlier we estimated a dust optical
depth of 0.25 at 1.4 mm, which would lead to us underestimating the
mass by only 10\% or so. On the other hand, the \hts\ is clearly
associated with outflow motions, which artifically raises the virial
mass (due to increased linewidths), leading to the observed large
discrepancy between the mass estimates. Using the (lower) mass
estimates from the dust emission to derive H$_2$ column densities will
lead us to derive higher values for the \hts\ abundance. Performing
this calculation yields \hts\ abundances a factor of 3--5 higher than
given in Table \ref{hparameters} (again dependent on the assumed dust
temperature).

The \hts\ virial masses for both component G9.62E and G9.62F are in
good agreement with the masses derived from \ceo\ by Hofner et al.\
(1996), and our \ceo\ virial mass agrees well with their total mass
for E and F. Since we measure a larger linewidth for our \hts\ data
(7.8 \kms\ compared with 4.4 \kms\ for ammonia), our estimate of the
virial mass for F is correspondingly higher than the Cesaroni et al.\
(1994) value although the sizes are in good agreement. Consequently,
we also derive higher densities. Our larger linewidth is a reflection
of the presence of the \hts\ in the outflow; ammonia is therefore not
present or excited in detectable quantities in the outflow.

Mass estimates for the other three sources vary considerably. Virial
masses derived from the ammonia (4,4) line (Cesaroni et al.\ 1994)
tend to be $\sim$5 times smaller than our values, largely due to the
higher \hts\ linewidth and larger extent in \ceo. Tracers employed at
3 mm by Maxia et al.\ (2001) and Olmi et al.\ (2003) yield estimates
similar to what we derive. Maxia et al.\ (2001) use a smaller dust
emissivity that we have employed; consequently we revise their
estimates derived from 3-mm dust continuum downwards by a factor of
$\sim$4 to give $\sim$700 M$_\odot$ and 350 M$_\odot$ for G29.96 and
G31.41 respectively. Olmi et al.\ (2003) estimate a mass of 1200
M$_\odot$ for G29.96 from \ceo, larger than our estimates (Table
\ref{cparameters}, \ref{hparameters}).

Submillimeter dust continuum models constructed by Hatchell et al.\
(2000) for G10.47 and G31.41 have a constant density core at the
center which has a mass of 3000 M$_\odot$ within a radius of 0.06 pc,
somewhat larger than our estimates for a similar-sized region.

The dimensions we derive for G9.62F are very similar to what Cesaroni
et al.\ (1994) determine from their ammonia data. In G10.47, G29.96
and G31.41 the size of the hot core derived from the \hts\ emission is
larger than that derived from the ammonia observations of Cesaroni et
al.\ (1994) and C98. C98 showed that the optically thicker main
hyperfine lines of ammonia yielded a larger source size than the
satellites. Since we find that the \hts\ emission is probably
optically thick, it seems plausible that the larger source size is due
to a high optical depth in the \hts\ lines. Alternatively, our BIMA
observations may be more sensitive to the distribution of \hts\
relative to the VLA observations of ammonia (Cesaroni et al.\ 1994). A
third (and perhaps most likely) possibility is that the ammonia (4,4)
line simply traces warmer gas than the \hts\ as the energy of the
upper level is 84\,K for the \hts\ while that of the NH$_3$ (4,4) line
is 201\,K.

\subsection{Outflow parameters}

In Table \ref{outflow} we list the derived properties of the outflows
toward each source, including the mass of \hts -emitting gas.  If we
assume an abundance, then we can estimate the mass of swept-up
material in the outflows from each source on the condition that the
\hts\ is well-mixed with the molecular hydrogen. The \hts\ abundance
is not well known in molecular clouds, and even less well known in
outflows. In the previous section, we derived column-averaged
abundances of order $10^{-9}$, which are probably lower limits. In
other work, Minh et al.\ (1991) derived values close to $10^{-8}$ in a
number of high-mass star-forming regions, and as high as $10^{-6}$ in
Orion (Minh et al.\ 1990).

For the gas in the outflow, we assume an abundance of $10^{-7}$
relative to molecular hydrogen as it lies in the middle of the range
of values derived by Minh et al.\ (1990, 1991), and $10^{-7}$ is used
as an initial value for the gas-phase \hts\ abundances in the majority
of chemical models (e.g. Rodgers \& Charnley 2003, Wakelam et al.\
2004). If the \hts\ abundance is as high as $10^{-6}$, then we derive
minimum outflow masses of 0.4 M$_\odot$, values typical of relatively
low-mass stars. On the other hand, if we assume smaller values then
the mass estimates will increase to very high values (40 M$_\odot$ per
lobe). Therefore, we regard our assumption of $10^{-7}$ for the \hts\
abundance as providing reasonable lower-limits to the outflow
parameters, although CO observations are desireable to determine these
parameters with greater certainty.

It appears that for all our sources several solar masses of material
has been swept up, typically $\sim$4 M$_\odot$ in each lobe. Since we
have neglected emission close to the line center, the true mass of
each outflow is probably higher. Even so, given the mass estimates in
Table \ref{outflow}, it is clear that these outflows are being driven
by massive protostars since they compare favorably with a number of
measurements of larger, more well-developed flows from massive YSOs
(e.g.\ Gibb et al.\ 2003; Shepherd et al.\ 1998).

If we assume that the outflow masses we derive are a sufficiently
accurate reflection of the true masses, we can also estimate other
properties of the outflows such as the momentum and energy. We assume
that the characteristic outflow velocity is $v_{\rm out} = v_{\rm
max}/2$ so that the momentum, $p$, is simply $mv_{\rm out}$, and the
energy is $E_{\rm K} = mv_{\rm out}^2/2$. Similarly, we assume the
outflow `size' ($r_{\rm out}$) to be half the mean diameter of the
outflow lobes. The dynamical `age', $t_{\rm dyn}$, is calculated from
$r_{\rm out}/v_{\rm out}$, with outflow force and mechanical
luminosity calculated as $p/t_{\rm dyn}$ and $E_{\rm K}/t_{\rm dyn}$.
Table \ref{out2} lists the values thus calculated. The average value
for each outflow is given; the total momentum, energy, etc., will be a
factor of two higher. All the values are consistent with the outflows
being driven by relatively massive YSOs (c.f. Shepherd \& Churchwell
1996), although the choice of \hts\ abundance clearly affects our
final numbers. However, we wish to point out that since we are not
sensitive to the highest-velocity gas (the velocity extent is quite
low) the momentum and energy for each outflow will be seriously
underestimated. High-$J$ CO observations are desirable in order to
better constrain the flow energetics.

\section{Discussion}

\hts\ has only been previously observed in single-dish observations of
these sources. Since we are able to resolve the structure of the \hts\
emission we can now discuss our observations in the context of
physical and chemical models of hot cores.

\subsection{\hts\ in hot cores: distribution and abundance}
\label{abundance}

Hatchell et al.\ (1998a) estimate the \hts\ column density toward all
four of our target sources from single-dish observations of the same
transition (and assuming the same excitation conditions) as we present
here. These single-dish column densities are typically a factor of 20
lower than the column densities we derive, due to the brightness
temperature being diluted by the 22-arcsec JCMT beam. Thus the filling
factor for the JCMT observations is $\sim$1/20, which predicts a
source size of order 5\arcsec, very similar to what we observe (Table
\ref{hparameters}).

In the absence of any other input, chemical models of hot cores start
with a gas-phase \hts\ abundance of around $10^{-7}$ based on the
observations of Minh et al.\ (1991), and more recently on the ISO
results which place an upper limit of $10^{-7}$ on the solid \hts\
abundance (van Dishoeck \& Blake 1996). For example, Rodgers \&
Charnley (2003) follow the evolution of chemical abundances in
centrally-heated cores assuming two different models for the evolution
of the physical parameters of the core. Rodgers \& Charnley suggest
that their static model is more appropriate for massive cores, and
show that the abundance of \hts\ is initially high and decreases
rapidly with time.  At early times ($< 10^4$ yr) \hts\ appears to have
a uniform abundance across the core, falling off at an outer radius
which increases with time, reaching 0.01 pc at $10^4$ yr (the outer
radius of their model is 0.03 pc). The typical abundance for \hts\ in
these models is a few $\times 10^{-8}$ to $10^{-7}$, falling off
rapidly with radius at late times (although it rises again in the
outer regions of the core).  Inspection of their predicted column
densities shows that only at late times (at $10^5$ yr and beyond) does
the \hts\ column density fall below $10^{16}$
cm$^{-2}$. Simultaneously, the size of the hot core in ammonia is at
least a factor of two smaller, purely as a result of the
chemistry. This appears to be another factor which can account for the
fact that these hot cores appear to be smaller in the ammonia images
of C98 than in our \hts.

We note that the hot core models predict \hts\ abundances at least an
order of magnitude greater than the values we calculate (Table
\ref{hparameters}). As discussed elsewhere in this paper, it is likely
that a combination of high optical depth and an overestimated mass
from the virial theorem can account for this discrepancy. Since
ammonia is well studied on similar scales in hot cores, we can examine
the NH$_3$/\hts\ abundance ratio. Using the ammonia column densities
given by Cesaroni et al.\ (1994) we derive ratios
[\hts/NH$_3$]=2--8$\times 10^{-3}$. However, in the models of Rodgers
\& Charnley (2003), this ratio never falls below unity except at late
times ($10^5$ yr). The simplest explanation for the observed ratios is
that the \hts\ lines have significant optical depth leading to vastly
underestimated \hts\ column densities.

\subsection{Hot core dynamics: outflow and rotation?}
\label{dynamics}

\begin{figure}
\figurenum{10}
\epsscale{0.8}
\plotone{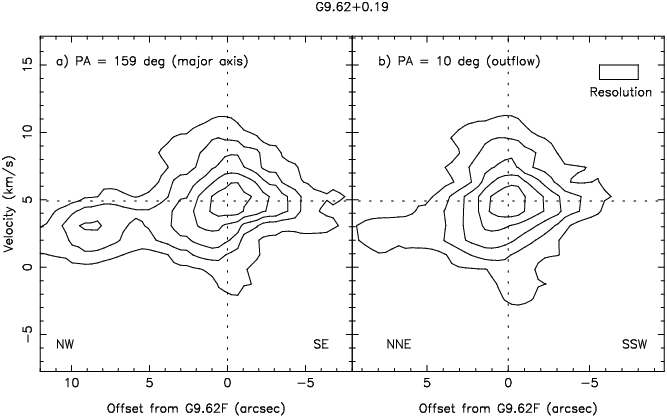}
\caption{G9.62 position-velocity diagrams along a) the core and b) the
  outflow axis respectively. The position angle of each cut is shown,
  measured positive east of north. The box in the top right-hand
  corner represents the effective resolution on each axis.}
\label{g9pv}
\end{figure}

\begin{figure}
\figurenum{11}
\epsscale{0.8}
\plotone{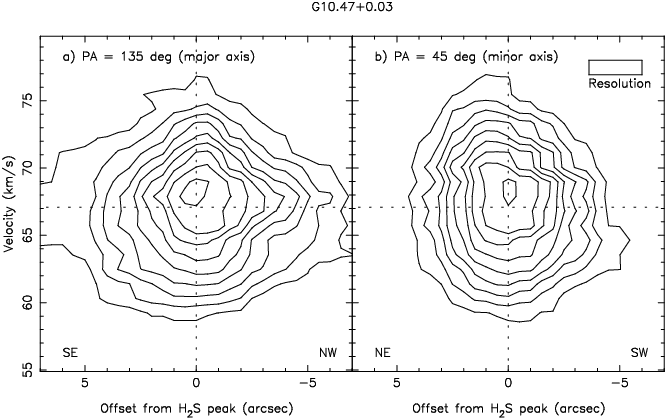}
\caption{G10.47. position-velocity diagrams along a) the major and b)
  the minor axes respectively. The position angle of each cut is
  shown, measured positive east of north. The box in the top
  right-hand corner represents the effective resolution on each axis.}
\label{g10pv}
\end{figure}

\begin{figure}
\figurenum{12}
\epsscale{0.8}
\plotone{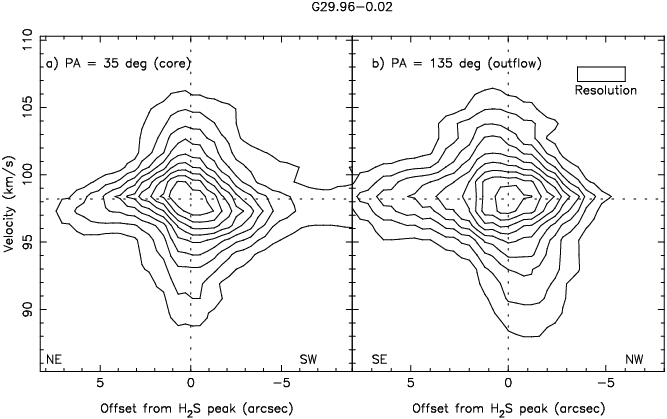}
\caption{G29.96. position-velocity diagrams along a) the core and b)
  the outflow axis respectively. The position angle of each cut is
  shown, measured positive east of north. The box in the top
  right-hand corner represents the effective resolution on each
  axis.}
\label{g29pv}
\end{figure}

\begin{figure}
\figurenum{13}
\epsscale{0.8}
\plotone{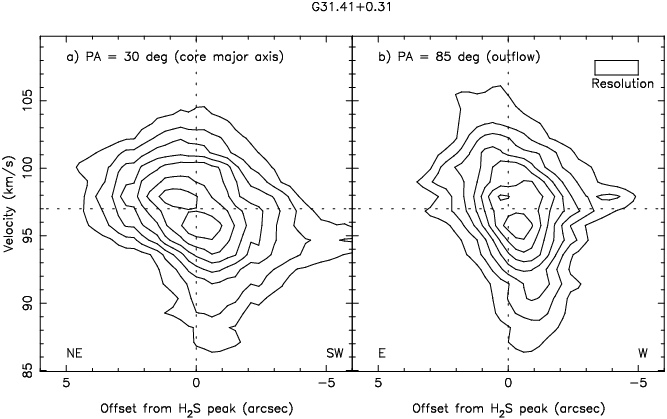}
\caption{G31.41. position-velocity diagrams along a) the core and b)
  the outflow axis respectively. The position angle of each cut is
  shown, measured positive east of north. The box in the top
  right-hand corner represents the effective resolution on each
  axis. }
\label{g31pv}
\end{figure}

The data we present here appear to trace two kinematic components. The
first is outflow traced by the \hts. The second is a velocity gradient
more or less perpendicular to the outflow and is probably rotation,
seen in both \ceo\ and \hts.  The outflow evidence is clear: spatially
bipolar lobes of red- and blueshifted gas. The rotation interpretation
is ambiguous, especially in light of the likely fragmented nature of
the G9.62 core and the more obvious outflow signature in the
\hts. However, given the good agreement between the \ceo\ and \hts\
gradients across each core, the rotation interpretation seems more
tenable. In Figures \ref{g9pv} to \ref{g31pv} we show selected
position-velocity (PV) cuts across each of the \hts\ data cubes for
each of the hot cores (along the long axis of the core and along the
outflow axis where defined). In three of the four sources clear
velocity gradients are present. Only G10.47 (Figure \ref{g10pv}) does
not show a velocity gradient. The observed velocity gradients tend to
be linear along the long axis of the core, although they typically
only extend for a couple of beamwidths. (However, we note that it is
possible to distinguish velocity gradients on scales smaller than a
beam: e.g. Olmi et al.\ 1996.) It is clear that the velocity gradients
do not directly show evidence for Keplerian rotation; indeed given the
massive cores within which the central stars are embedded such a
result would be surprising since a Keplerian rotation law only
strictly applies for a dominant point mass (although it may be true
that the velocity field is locally Keplerian). Furthermore, if the
observed emission is optically thick (as concluded above) then
regardless of the intrinsic velocity field, the observed velocity
gradient will be linear.

However, even if the emission is not optically thick, our finite
beamsize prevents us from directly observing a Keplerian rotation
curve, instead smoothing out the central region into a linear gradient
(see Richer \& Padman 1991).  Therefore, we caution against
over-interpreting these PV data, especially since the emission closest
to the center is heavily contaminated by the high-velocity emission
from the outflow. At best we can only conclude that in the cases of
G9.62, G29.96 and G31.41 our \ceo\ and \hts\ data show evidence for
velocity gradients along the long axis of the hot core, but not what
form the velocity law takes.

If we assume the linear velocity gradients are due to uniform rotation
then we can estimate rotation periods, and then the masses required to
bind the motions. For each of the three sources with clear velocity
gradients (G9.62F, G29.96 and G31.41), the rotation periods for each
core is similar at 1--2$\times 10^5$ yr. Using the core dimensions
shown in Table \ref{hparameters} these periods require binding masses
of only 15, 60 and 154 M$_\odot$ (for G9.62F, G29.96 and G31.41
respectively). These masses are all less than both the virial and
dust-derived masses and so that the cores are rotationally bound, and
may even be still collapsing. The velocity gradients along the outflow
axes are larger and imply binding masses of 810, 2800 and 2360
M$_\odot$ respectively, all much larger than the virial masses given
in Tables \ref{cparameters} and \ref{hparameters}. Therefore the
motions along the outflow axes are not bound to the hot cores.

An interesting feature of the G31.41 PV diagram is that the \hts\
emission farthest from the center does not lie at the rest
velocity. This indicates that the \hts\ is not very abundant at
these distances from the hot core and falls below our detection
threshold. From Figure \ref{g31pv} we can therefore determine an outer
radius of the hot core of $\sim$5 arcsec or 0.19 pc ($5.9\times
10^{17}$ cm).

\subsubsection{An outflow origin for methyl cyanide?}
\label{ch3cn}

Comparing our results with those of Beltr\'an et al.\ (2004) for
G31.41, we find that our outflow direction is closely aligned with the
long axis of their methyl cyanide emission, which they interpret as a
rotating disk or toroid. Their interpretation was a natural conclusion
given the example of IRAS\,20126+4104 (Cesaroni et al.\ 1999) which
shows a clear disk-like component for the CH$_3$CN emission. This
ambiguity highlights the kind of problems faced when studying hot
cores.  However, we believe that our \ceo\ and \hts\ data display
clear evidence for both outflow and (in two cases) velocity gradients
perpendicular (or close to) to the axis of the outflow. The simplest
interpretation for these gradients is rotation of the core.
The distribution of our \hts\ (and 1.4-mm continuum) data closely
matches that of the NH$_3$ (4,4) of C98. This leads to the interesting
conclusion that Beltr\'an et al.\ (2004) have actually detected
CH$_3$CN in the outflow from G31.41. Further support comes from the
CH$_3$CN observations of G29.96 by Olmi et al.\ (2003) which are also
consistent with emission from an outflow since the velocity gradient
is extremely high (higher than we detect with our \hts\ data, thus
requiring an even higher binding mass) and oriented mostly along the
same direction as our outflow.  It is interesting to compare the
methyl cyanide and \hts\ results. \hts\ is clearly not confined to the
outflow. On the other hand, it seems as if the CH$_3$CN in these
sources is dominated by an outflow component, suggesting that it is
perhaps shock chemistry that is playing a greater role in the
formation of methyl cyanide. Clearly further high-resolution
observations are necessary in order to explore this further.

\subsection{Embedded sources in hot cores}

The primary result from these observations is the confirmation that
outflows emanate from some hot cores, including all four of the
targets in our sample. Furthermore the images of the high-velocity
emission show that these outflows appear to be collimated; in other
words, unlike many ultracompact \HII\ regions, they are not spherical
shells driven by the expansion of a symmetric stellar wind. This has
important implications for the formation of massive stars and implies
the presence of a collimation mechanism. In the case of low-mass
stars, this is now widely believed to be a rotating circumstellar disk
(Ouyed, Clarke \& Pudritz 2003; K\"onigl \& Pudritz
2000). Explanations of outflows from more luminous protostars invoke a
wider-angle contribution from a stellar wind (Shepherd et al.\ 1998;
Richer et al.\ 2000), although a disk is still believed to play an
important role in at least initially collimating the outflow.

Clearly we cannot determine whether circumstellar disks exist around
these outflow sources with the current observations. However, we can
examine whether the ambient medium has sufficient turbulent pressure
to act as a collimating agent for the flows. Assuming a single driving
source we may equate the momentum flux from the wind/outflow (assumed
to be initially isotropic) with the turbulent back-pressure of the
ambient medium:
\begin{equation}
\rho \sigma_{\rm 1D}^2 = \frac{\dot{p}}{4\pi r^2}
\end{equation}
where $\rho$ is the ambient density, $\sigma_{\rm 1D}^2$ is the
1-dimensional velocity dispersion ($=\sqrt{8 \ln 2} \Delta v$, where
$\Delta v$ is the FWHM), $\dot{p}$ is the outflow force and $r$ is the
radius at which the forces are equal (the collimation radius). Using
values tabulated in Tables \ref{hparameters} and \ref{outflow} we
derive collimation radii of 350 to 950 AU. At the distances to our
sources, these radii correspond to angular scales of 0.06\arcsec to
0.13\arcsec, and are thus unresolvable with current facilities. If
multiple sources contribute to the flow then the expression above can
be applied for each source assuming a momentum flux equal to the
appropriate fraction of the total. This will lead to smaller
collimation radii and our conclusion is unchanged. Therefore, it seems
likely that the ambient medium is playing a significant role in
producing the bipolar appearance of the outflows from these hot
cores. The fact that outflows are observed clearly points to a
non-spherical geometry.

Another implication from our observations is that the outflow must
begin at a stage prior to the formation of an ultracompact \HII\
region since the hot cores in G29.96 and G31.41 have no radio sources
within them. It could be argued that G10.47 provides evidence counter
to this assertion, but we note that C98 conclude the \HII\ regions in
G10.47 are located toward the {\em front} of the hot core, and are
therefore not deeply embedded within the hot core itself. However,
G10.47 is our weakest example of an outflow, and the hot core may only
be a hot, dense remnant of the formation of the stars exciting the
\HII\ regions, as Watt \& Mundy (1999) concluded for
G34.26. Unfortunately it is not possible to assess the likelihood of
this interpretation with our data and further observations are
desirable.

G9.62F has a radio source within it, but it is very weak, and probably
represents an ionized stellar wind rather than an \HII\ region (Testi
et al.\ 2000). The Orion hot core is probably excited by the radio
source I (Menten \& Reid 1995), which has a 22-GHz flux density of 4.8
mJy (Plambeck et al.\ 1995). If source I were located at a distance
typical of the sources in this paper ($\sim$6.5 kpc) then it would be
(currently) undetectable at 22 GHz (the 22-GHz flux density would only
be 30 $\mu$Jy). Equally, BN would also not be detectable. Therefore it
is possible that equivalent sources are located within the hot cores
studied here. It should be noted that the centimeter radio emission in
both BN and source I is from an ionized stellar wind rather than an
ultracompact \HII\ region.

Could it be that there are \HII\ regions in G29.96 and G31.41 but they
are not detectable due to their greater distance? We can test this by
scaling the fluxes of the \HII\ regions in G10.47, and using the
values given in C98 for the 1.3-cm flux, we find that such \HII\
regions would be detectable at the distance of G29.96 and
G31.41. Therefore the non-detection of \HII\ regions toward these hot
cores suggests that ultracompact \HII\ regions have not yet formed
within them, although stellar wind sources may be present.  Since the
outflow phenomenon is intimately linked to accretion, our observations
are also compatible with the hypothesis that \HII\ regions can be
quenched by a strong accretion flow (Walmsley 1995).

Another interesting point to note is that in each case, there appears
to be only a single bipolar outflow detected. (If the observed bipolar
outflow is the superposition of multiple flows, then they clearly
share a common orientation.) This is surprising given the clustered
nature of massive star formation: it seems more likely that we would
not detect an orderly bipolar flow from such a crowded
environment. Indeed, even in binary systems the outflow orientations
can be distinctly different: the low-mass systems in HH111 and
NGC\,1333-IRAS2 both show pairs of outflows with very different
position angles (Cernicharo \& Reipurth 1996; Knee \& Sandell 2000),
while IRAS\,16293$-$2422 exhibits a quadrapolar flow (Walker,
Carlstrom \& Bieging 1993). The observation of a single flow has
important implications for the formation of massive stars. One
possiblity is that the most massive and energetic flow (presumably
from the most massive YSO) dominates the energetics, in a similar
fashion to how infrared and radio continuum observations are dominated
by the most luminous source. Alternatively, the outflow stage could be
so short that at any one time, we only catch one source in the act. An
extension of this interpretation is that we have detected these
outflows in \hts, a molecule which is itself comparatively short-lived
in hot cores (Charnley 1997). It is possible that the short outflow
lifetime coupled with the short chemical lifetime for \hts\ reinforce
each other to produce a single outflow detection. Of course it is
highly likely that multiple sources are present within these hot
cores, each potentially with its own outflow. High-resolution
observations of CO, whose abudance is well-known and stable, are
needed to search for multiple outflows.

As always it is instructive to compare these results with the nearest
hot core in Orion. The outflow from the Orion hot core is well-defined
in CO (although the collimation is not very high; e.g. Chernin \&
Wright 1995) but has not been probed in \hts. The CO 4\too 3 data of
Schulz et al.\ (1995) show a bipolar outflow with outflow lobes
separated by $\sim$20\arcsec. If such a flow were moved ten to twenty
times further away then it is the CO flow would only be barely
resolvable with current interferometers.

\subsection{Scattered light at mid-IR wavelengths?}

An interesting feature of our G29.96 results is that the 18 $\mu$m
peak of De Buizer et al.\ (2002) is offset to the NW of the 1.4-mm
continuum peak (Figure \ref{g2996}c). It lies within the blue lobe
close to the position of the methanol masers observed by Minier et
al.\ (2002). This leads to the hypothesis that even at 18 $\mu$m we
are seeing scattered light. If this is the case, then we expect the 10
$\mu$m emission to peak further from the outflow center.  This does
not appear to be the case but it is difficult to be sure since the hot
core is weaker relative to the UC\HII\ region at 10 $\mu$m.  This
interpretation is tentative at best since the warm dust in the \HII\
region dominates the emission at these mid-IR wavelengths and makes
the isolation of the hot-core emission non-trivial. However, it is
useful to examine our data for further support.

In Section \ref{masses} above, we estimated that the extinction toward
the hot cores is significant even in the mid-infrared (450 to 520
magnitudes at 20 $\mu$m). This, and the extra extinction caused by the
18-$\mu$m silicate absorption feature (e.g.\ Mathis 1990), would
result in the hot core emission to be strongly self-absorbed at
18-$\mu$m. Therefore the only way that light can escape is through a
cavity. An outflow provides a natural explanation for the presence of
a cavity, and so we conclude that the evidence strongly suggests that
the 18-$\mu$m source detected by De Buizer et al.\ (2002) is in fact
scattered light. It is interesting to note the recent conclusion of
McCabe, Duch\^ene \& Ghez (2003) that the 12-$\mu$m emission toward
the T Tauri star HK Tau B appears to be scattered light.

Hofner et al.\ (2001) argue that the outflow in G9.62F is observed
almost pole-on since weak near-infrared emission is detected. If our
scattered-light interpretation is correct, then G9.62F should be
easily detected at 18-$\mu$m and observed coincident with the hot
core, since the dust column to the central source will be much
less. Similar observations of G10.47 and G31.41 are desirable to test
whether these sources are visible at 18 $\mu$m.


\section{Conclusions: collimated outflows from rotating hot cores}

We have presented high-resolution observations made with the BIMA
millimetre array of the 2(2,0)\too 2(1,1) transition of \hts\ and
2\too 1 transition of \ceo\ toward a sample of four hot cores:
G9.62+0.19, G10.47+0.03, G29.96$-$0.02 and G31.41+0.31. In three of
the four cases (G9.62F, G29.96 and G31.41), high-velocity \hts\
emission is observed which is distributed in a spatially-bipolar
fashion, in a direction offset from the major axis of the hot core (as
defined by the integrated intensity). In addition, linear velocity
gradients are observed in both \ceo\ and \hts\ along the major axis of
these three cores. Only in G10.47+0.03 are the data ambiguous, showing
neither clear rotation nor outflow signatures.

Although the \hts\ abundance is not well known, we have made
reasonable assumptions to derive the parameters of these
outflows. These estimates show that the outflows are indeed driven by
massive protostars despite being subject to large uncertainties in the
\hts\ abundance.

The lack of centimeter-wave radio emission from within two of the four
hot cores shows that outflows begin at a phase prior to the generation
of an ultracompact \HII\ region. Furthermore, our data show only one
outflow toward each hot core. This implies that the observed outflow
in \hts\ is dominated by a single source, either by the most massive
or by the source which we just happen to catch in this particular
phase of evolution.

We interpret these data as indicating the presence of a massive
protostar embedded within a hot core, which may be rotating, and
driving a molecular outflow.  The bipolar nature implies that these
flows are collimated, and we have shown that the ambient material is
capable of channeling flows with the properties derived here and, if
this is the primary collimation mechanism, then the mass distribution
is not spherically symmetric.

\acknowledgments
This work was supported by the National Science Foundation under
grants AST-9613716, AST-9981289 and AST-0028963. The referee is
thanked for their comments and suggestions.


\bibliography{}
\bibliographystyle{astron}

\clearpage


\begin{deluxetable}{lccccccccc}
\tablewidth{0pt}
\tabletypesize{\footnotesize}  

\tablecaption{Source details, coordinates for pointing centers and beam parameters.
\label{sources} }

\tablehead{
Source & $\alpha$(J2000)\tablenotemark{a} & $\delta$(J2000)\tablenotemark{b} & $d$ & $v_{\rm LSR}$ & $L_{\rm bol}$ &
\multicolumn{2}{c}{\hts} & \multicolumn{2}{c}{\ceo} \\
       &               & &          &          &             & Beam & PA & Beam & PA \\
       &               & &    (kpc) & (\kms)   & (L$_\odot$) & (arcsec$^2$)  & (deg) & (arcsec$^2$)  & (deg)
}
\startdata
\objectname{G9.62+0.19}    & 18:06:14.812 & $-$20:31:39.390 & 5.7 & \phn4.4 & 4.4(5) &  3.4$\times$1.7 & \phn2.5  & 4.5$\times$2.3 & \phn6.7 \\
\objectname{G10.47+0.03}   & 18:08:38.280 & $-$19:51:50.000 & 5.8 & 67.8    & 5.0(5) &  2.8$\times$1.3 & \phn9.2  & 3.5$\times$1.9 & \phn8.6 \\
\objectname{G29.96$-$0.02} & 18:46:03.775 & $-$02:39:21.884 & 7.4 & 97.6    & 1.4(6) &  2.0$\times$1.3 & \phn7.0  & 3.5$\times$1.9 & \phn8.4 \\
\objectname{G31.41+0.31}   & 18:47:34.320 & $-$01:12:45.800 & 7.9 & 97.4    & 1.7(5) &  1.8$\times$1.2 &    24.8  & 2.4$\times$1.7 &    24.5 \\
\enddata
\tablecomments{Bolometric luminosities are taken
  from Cesaroni et al.\ (1994) and Hatchell et al.\ (2000). The
  notation $a(b)$ represents $a \times 10^b$. } 
\tablenotetext{a}{In hours, minutes and seconds.}
\tablenotetext{b}{In degrees, minutes of arc and seconds of arc. }
\end{deluxetable}


\begin{deluxetable}{lcccccccc}
\tablewidth{0pt}

\tablecaption{Observed properties of each source \ceo\ and \hts. \label{properties}}

\tablehead{
Source & \multicolumn{4}{c}{\ceo} & \multicolumn{4}{c}{\hts}  \\
 & \multicolumn{2}{c}{FWHM Dimensions\tablenotemark{a}} & PA &
$\langle \Delta v\rangle$\tablenotemark{b} & \multicolumn{2}{c}{FWHM
  Dimensions\tablenotemark{a}} & PA & $\langle\Delta
v\rangle$\tablenotemark{b}  \\
 & (arcsec$^2$) & (pc$^2$) & (deg) & (\kms) & (arcsec$^2$) & (pc$^2$) & (deg) & (\kms) }

\startdata
G9.62+0.19    & 11.4$\times$7.0 & 0.32$\times$0.19 & $-$33 & 4.2$\pm$1.2 &                &                  &       &             \\
G9.62+0.19E   &                 &                  &       &             & 2.9$\times$2.0 & 0.08$\times$0.06 & $-$29 & 4.1$\pm$0.9 \\
G9.62+0.19F   &                 &                  &       &             & 4.6$\times$2.2 & 0.12$\times$0.06 & $-$24 & 5.3$\pm$1.3 \\
G10.47+0.03   & 6.4$\times$5.2  & 0.18$\times$0.15 & $-$42 & 7.4$\pm$1.8 & 5.5$\times$3.6 & 0.15$\times$0.10 & $-$31 & 8.2$\pm$1.7 \\
G29.96$-$0.02 & 9.1$\times$6.8  & 0.33$\times$0.24 & $-$70 & 3.2$\pm$0.5 & 4.3$\times$3.2 & 0.16$\times$0.12 & $-$18 & 5.2$\pm$1.9 \\
G31.41+0.31   & 12.7$\times$6.1 & 0.49$\times$0.23 &  +22  & 4.5$\pm$1.6 & 3.5$\times$2.5 & 0.14$\times$0.10 &  +26  & 6.6$\pm$1.8 \\
\enddata

\tablecomments{The notation $a(b)$ represents $a \times 10^b$.}
\tablenotetext{a}{Deconvolved dimensions obtained from a two-dimensional
gaussian fit to the integrated intensity image. The uncertainty is
typically less than 0.25$''$, or less than 0.01 pc at the distance to
all sources.}
\tablenotetext{b}{Mean linewidth within the half-power
  contour. Uncertainty represents dispersion in values.}
\end{deluxetable}

\begin{deluxetable}{lccccc}
\tablewidth{0pt}

\tablecaption{Parameters derived from \ceo\  data. \label{cparameters}}

\tablehead{
Source & $M_{\rm
  vir}$\tablenotemark{a} & $N_{\rm H_2}$\tablenotemark{a} & $n_{\rm H_2}$\tablenotemark{a} & $N_{\rm C^{18}O}$\tablenotemark{b,{\rm c}} & $M_{\rm
  H_2}$\tablenotemark{b,{\rm d}} \\
 & (M$_\odot$) & (cm$^{-2}$) & (cm$^{-3}$) & (cm$^{-2}$) & (M$_\odot$)   }

\startdata
G9.62+0.19    & 463$\pm$137    & 7.7$\pm$2.3(23) &  1.0$\pm$0.3(6) & 1.2$\pm$0.6(17) & 303$\pm$141 \\
G10.47+0.03   & 920$\pm$252    & 3.8$\pm$1.0(24) &  7.7$\pm$2.1(6) & 1.8$\pm$0.8(17) & 172$\pm$\phn80 \\
G29.96$-$0.02 & 301$\pm$\phn52 & 4.0$\pm$0.7(23) &  4.6$\pm$0.8(5) & 9.7$\pm$4.5(16) & 346$\pm$161 \\
G31.41+0.31   & 723$\pm$261    & 6.5$\pm$2.3(23) &  6.2$\pm$2.2(6) & 1.5$\pm$0.7(16) & 264$\pm$123 \\
\enddata

\tablecomments{The notation $a(b)$ represents $a \times
  10^b$. }
\tablenotetext{a}{Calculated assuming a uniform sphere.}
\tablenotetext{b}{Calculated assuming optically thin emission and an
  excitation temperature of 75 K. Uncertainty reflects excitation
  temperatures in the range 16 to 125 K.} 
\tablenotetext{c}{At the hot core position }
\tablenotetext{d}{Mass calculated within the half-power contour of the
  integrated \ceo\ emission, assuming a \ceo\ abundance of 2$\times$10$^{-7}$.}
\end{deluxetable}

\begin{deluxetable}{lccccc}
\tablewidth{0pt}

\tablecaption{Parameters derived from \hts\ data. \label{hparameters}}

\tablehead{
Source &  $M_{\rm
  vir}$\tablenotemark{a} & $N_{\rm H_2}$\tablenotemark{a} & $n_{\rm H_2}$\tablenotemark{a} & $N_{\rm H_2S}$\tablenotemark{b} & $X_{\rm H_2S}$ \\
 &  (M$_\odot$) & (cm$^{-2}$) & (cm$^{-3}$) & (cm$^{-2}$) &  }

\startdata
G9.62+0.19E   & 124$\pm$\phn45 & 2.6$\pm$1.0(24) & 1.2$\pm$0.4(7) & $>$3.0(15) & $>$1.2($-$9) \\
G9.62+0.19F   & 248$\pm$\phn85 & 3.7$\pm$1.3(24) & 1.4$\pm$0.5(7) & $>$1.2(16) & $>$3.3($-$9) \\
G10.47+0.03   & 861$\pm$228    & 6.0$\pm$1.6(24) & 1.6$\pm$0.4(7) & $>$4.1(16) & $>$6.8($-$9) \\
G29.96$-$0.02 & 392$\pm$154    & 2.2$\pm$0.8(24) & 5.0$\pm$2.0(6) & $>$1.3(16) & $>$6.1($-$9) \\
G31.41+0.31   & 540$\pm$173    & 4.1$\pm$1.3(24) & 1.1$\pm$0.4(7) & $>$2.7(16) & $>$6.7($-$9) \\
\enddata

\tablecomments{The notation $a(b)$ represents $a \times
  10^b$. }
\tablenotetext{a}{Calculated assuming a uniform sphere.}
\tablenotetext{b}{Calculated at the peak position assuming optically thin emission and an
  excitation temperature of 56 K.} 
\end{deluxetable}


\begin{deluxetable}{lccccccc}
\tablewidth{0pt}
\tablecaption{Outflow properties. \label{outflow}} 

\tablehead{Source & \multicolumn{3}{c}{Red lobe} & \multicolumn{3}{c}{Blue lobe}  & \\
       & Size\tablenotemark{a}     & Velocity\tablenotemark{b} & $M_{\rm H_2S}$    & Size\tablenotemark{a} & Velocity\tablenotemark{b} & $M_{\rm H_2S}$ & \\
       & (pc$^2$) & (\kms)   & (M$_\odot$)  & (pc$^2$) & (\kms)   & (M$_\odot$)  & 
}
\startdata
G9.62+0.19F   & 0.12$\times$0.07 & \phn+6.5 & 4.4($-$7) & 0.14$\times$0.08 & \phn$-$8.5 & 4.0($-$7) & \\
G10.47+0.03   & 0.13$\times$0.07 & \phn+7.5 & 3.5($-$7) & 0.14$\times$0.08 &    $-$10.0 & 5.4($-$7) & \\
G29.96$-$0.02 & 0.13$\times$0.08 & \phn+7.5 & 4.5($-$7) & 0.15$\times$0.09 &    $-$12.2 & 4.0($-$7) & \\
G31.41+0.31   & 0.11$\times$0.07 &    +11.9 & 5.0($-$7) & 0.07$\times$0.04 &    $-$13.8 & 3.7($-$7) & \\
\enddata

\tablecomments{The notation $a(b)$ represents $a \times 10^b$.}
\tablenotetext{a}{Deconvolved FWHM dimensions derived from a two-dimensional gaussian fit. } 
\tablenotetext{b}{The maximum flow velocity relative to the source systemic velocity. }

\end{deluxetable}

\begin{deluxetable}{lccccccccc}
\tabletypesize{\scriptsize}
\tablewidth{0pt}

\tablecaption{Estimated outflow energetics\tablenotemark{a}. \label{out2}}

\tablehead{
Source & $r_{\rm out}$ & $v_{\rm out}$ & $t_{\rm dyn}$ & $M$         & $\dot{M}$ 
& $p$               & $E_{\rm K}$ & $\dot{p}$                        & $L_{\rm mech}$ \\
       &   (pc)        & (\kms)        &   (yr)        & (M$_\odot$) & (M$_\odot$\,yr$^{-1}$) 
& (M$_\odot$\,\kms) & (erg)       & (M$_\odot$\,\kms \,yr$^{-1}$)   & (L$_\odot$)
}
\startdata
G9.62+0.19F   & 0.10 & 3.8 & 2.6(4) & 4.2 & 1.6($-$4) & 16 & 6.1(44) & 6.2($-$4) & 0.2 \\
G10.47+0.03   & 0.10 & 4.4 & 2.2(4) & 4.4 & 2.0($-$4) & 19 & 8.5(44) & 8.8($-$4) & 0.3 \\
G29.96$-$0.02 & 0.11 & 4.9 & 2.2(4) & 4.2 & 1.9($-$4) & 21 & 1.0(45) & 9.4($-$4) & 0.4 \\
G31.41+0.31   & 0.07 & 6.4 & 1.1(4) & 4.4 & 4.0($-$4) & 28 & 1.8(45) & 2.6($-$3) & 1.3 \\
\enddata
\tablecomments{The notation $a(b)$ represents $a \times 10^b$.}
\tablecomments{Values listed are the average of the those for each
  lobe of the outflow. Totals will be a factor of two greater. }
\tablenotetext{a}{An \hts\ abundance of $10^{-7}$ relative to H$_2$ has
  been assumed.}

\end{deluxetable}

\clearpage

\end{document}